\documentclass[a4paper]{article}
\usepackage{graphicx}
\usepackage{amsmath}
\usepackage{amsfonts}
\usepackage{amssymb}
\graphicspath{{images/}}
\usepackage{indentfirst}
\usepackage{caption}
\usepackage{float}
\usepackage{booktabs}
\usepackage{changepage}
\usepackage{array} 
\usepackage[english]{babel} 
\usepackage{hyperref} 
\usepackage{authblk}
\usepackage{geometry}
\title{Hybrid Models for Financial Forecasting: Combining Econometric, Machine Learning, and Deep Learning Models}
\date{} 

\author[1,*]{Dominik Stempień}
\author[2]{Robert Ślepaczuk}

\affil[1]{\small Faculty of Economic Sciences, University of Warsaw, Długa 44/50, 00-241 Warsaw, Poland}

\affil[2]{\small Department of Quantitative Finance and Machine Learning, Faculty of Economic Sciences, University of Warsaw, Długa 44/50, 00-241 Warsaw, Poland}

\affil[*]{\textbf{Corresponding author(s).} E-mail(s): \texttt{d.stempien@student.uw.edu.pl}}

\begin{document}

\maketitle

\begin{abstract}

This research systematically develops and evaluates various hybrid modeling approaches by combining traditional econometric models (ARIMA and ARFIMA models) with machine learning and deep learning techniques (SVM, XGBoost, and LSTM models) to forecast financial time series. The empirical analysis is based on two distinct financial assets: the S\&P 500 index and Bitcoin. By incorporating over two decades of daily data for the S\&P 500 and almost ten years of Bitcoin data, the study provides a comprehensive evaluation of forecasting methodologies across different market conditions and periods of financial distress. Models' training and hyperparameter tuning procedure is performed using a novel three-fold dynamic cross-validation method. The applicability of applied models is evaluated using both forecast error metrics and trading performance indicators. The obtained findings indicate that the proper construction process of hybrid models plays a crucial role in developing profitable trading strategies, outperforming their individual components and the benchmark Buy\&Hold strategy. The most effective hybrid model architecture was achieved by combining the econometric ARIMA model with either SVM or LSTM, under the assumption of a non-additive relationship between the linear and nonlinear components.

\bigskip
\noindent\textbf{\textit{Keywords:}} \textit{trading, quantitative finance, ARIMA, ARFIMA, Support Vector Machines, XGBoost, Long Short-Term Memory, hybrid models, walk-forward optimization, risk-adjusted metrics, portfolio}

\bigskip
\noindent\textbf{\textit{JEL Codes:}} C4, C14, C22, C45, C53, C58, G11, G17

\end{abstract}

\section{Introduction}

Financial time series analysis is crucial in making investment decisions, assessing market risks, and developing robust trading strategies. Despite the Efficient Market Hypothesis (Fama, 1970) and later research supporting the thesis (Malkiel, 2003), there is a broad body of studies focused on applying techniques to predict future market movements (Hsieh et al., 2011; Hsu et al., 2016; Weng et al., 2017). However, owing to the low signal-to-noise ratio and inherently chaotic nature of market data, the task of financial time series forecasting requires the development of a complex and detailed methodology appropriate to these challenges (De Prado, 2015). An effective application of predictive models may facilitate the construction of profitable investment strategies characterized by high profits and relative robustness to market fluctuations (Atsalakis and Valavanis, 2009). For this reason, the implementation of novel model architectures for predicting market movements constitutes an area of interest not only to researchers but also to individual investors and market practitioners. 

The traditional approach to time series forecasting is represented by a suite of various econometric models. This group of methods usually adopts assumptions concerning the linear, stationary, and normal distribution properties present in the data (Shah et al. 2019). The adoption of statistical techniques has found applications in many areas of financial forecasting for various classes of assets (Koustas and Serletis, 2005; Bhardwaj and Swanson, 2006; Kumar, 2010). However, with the development of more advanced machine learning models, the econometric approaches started to diminish in importance (Pérez-Pons et al., 2022). Machine learning techniques have allowed for exploring complex relationships within the data, including nonlinear patterns, without relying on potentially unrealistic assumptions about the underlying data-generating process (De Prado, 2019). As a result, there has been a rapid growth in financial research employing various machine learning and deep learning methods, distinguished by the use of different types of input data (Bustos and Pomares-Quimbaya, 2020; Rouf et al., 2021). Another forecasting methodology is based on combining predictions of individual econometric and machine learning models. The hybridization relies on the assumption that these models will manage to sequentially extract both the linear and nonlinear patterns in the data, which proved to be effective in improving individual forecasts (Aladag et al., 2012). Consequently, hybrid methods constitute an alternative to machine learning in modeling the nonlinear data, thus finding applications in the field of financial prediction (Rather et al., 2015). 

This study systematically develops and evaluates various hybrid modeling approaches by combining traditional econometric models with machine learning and deep learning techniques to forecast financial time series. The econometric segment is represented by the ARIMA and ARFIMA models, while the machine learning component comprises three distinct supervised learning techniques: SVM, XGBoost, and LSTM. The hybridization process is performed using forecasts generated by the constructed individual models, based on two alternative methodologies. The first approach, derived by Zhang (2003), consists of inputting the residuals from statistical models into machine learning models and then summing the resulting predictions. The second approach does not assume an additive relationship between linear and nonlinear components. Instead, the predictions from the econometric model are directly used as an additional explanatory feature in the machine learning model. As a result, the set of applied models in this study, including both individual and hybrid techniques, is composed of 17 distinct predictive methods. This complex framework allows for a thorough analysis of the applicability of hybridization in the field of financial forecasting.

The empirical investigation is based on two distinct financial assets: the S\&P 500 stock market index and Bitcoin. For S\&P 500, the data period covers daily observations from 1 January 2002 to 31 December 2023. In the case of Bitcoin, the data was collected throughout the period from 1 January 2015 and 31 December 2023. By incorporating over two decades of daily data for the S\&P 500 index and almost ten years of Bitcoin data, the study provides a comprehensive evaluation of forecasting methodologies across different market conditions. Especially, the market distress periods of the 2008 global financial crisis and the outbreak of the COVID-19 pandemic in 2020 are included. The applied models are trained and evaluated over the adopted time frames using cross-validation on a rolling basis with dynamically adjusted windows. For this purpose, an innovative time series validation technique, inspired by Choi et al. (2024), is proposed in this study. The accuracy and quality of the predictions for each of the constructed models are evaluated using standard error metrics: Root Mean Squared Error (RMSE) and Mean Absolute Error (MAE). Furthermore, the generated forecasts are transformed into two categories of trading signals. This stage enables the construction of corresponding trading strategies and the backtesting of the models on the real-world historical market data. The performance of the obtained investment strategies is examined based on the following trading metrics: annualized rate of return (ARC), annualized standard deviation (ASD), maximum drawdown (MD), information ratio (IR), adjusted information ratio (IR*), and Sortino ratio (SR). This set of measures facilitates a detailed analysis of various aspects of the investment outcomes. The inclusion of both forecast error metrics and trading performance indicators provides a more comprehensive assessment of the individual and hybrid models' effectiveness.

The primary objective of this research is to find data-based answers to the following questions:
\begin{itemize}
\item RQ1: \textit{Does the use of hybrid models yield more accurate predictions compared to individual linear models and machine learning techniques?}
\item RQ2: \textit{Is it possible to create a profitable trading strategy (compared to a buy-and-hold benchmark) using the forecasts of constructed hybrid models?}
\item RQ3: \textit{Which econometric model is better suited for hybrid models in describing linear dependencies in financial markets: the Autoregressive Integrated Moving Average (ARIMA) model or the Autoregressive Fractionally Integrated Moving Average (ARFIMA) model?}
\item RQ4: \textit{Which machine learning or deep learning model is better suited for hybrid models in describing non-linear dependencies in financial markets: Long Short-Term Memory (LSTM), eXtreme Gradient Boosting (XGBoost), or Support Vector Machines (SVM)?}
\item RQ5: \textit{Does the selection of hybridization technique influence the results?}
\item RQ6: \textit{Does the selection of the best hybrid model depend on the financial asset class?}
\end{itemize}

This study stands out from the existing literature on the subject in several ways. Firstly, a thorough comparative analysis of a series of hybrid models derived from various statistical and machine learning methodologies is undertaken. To achieve this purpose, the aim is to identify the combination of methods that efficiently capture both linear and nonlinear patterns in the data. By comparing hybrid models with the results obtained from traditional or machine learning models, the goal is to decide whether the application of hybrid methodology in financial time series forecasting is justified. Furthermore, different techniques of combining models are evaluated to check whether the method of hybridization influences the results. The objective is to ensure a deep understanding of hybridization methodologies, with a focus on each stage of the process - from selecting the underlying individual models to applying the most effective technique of merging their predictions. Secondly, the performance of the constructed models is assessed across multiple datasets to decide whether the characteristics of different financial assets influence the selection and performance of chosen models. In addition, a novel time series validation method is employed to ensure robust model evaluation. The aim is to provide a holistic understanding of hybrid models' performance across diverse financial markets while also evaluating the robustness across different market conditions. Lastly, the analysis covers not only the accuracy of models’ predictions but also the profitability of trading strategies based on these forecasts. That approach ensures the practical application of the findings to real-world financial markets with transaction costs. As a result, the project contributes valuable insights to academic researchers and individual traders. It empowers researchers with an enhanced understanding of hybrid modeling and its applicability in financial forecasting. Simultaneously, it equips traders with a structured framework to develop data-driven trading strategies, enabling them to perform more informed decision-making.

Most of the calculations in this study were performed using the Python programming language. The exception was made for the parameter estimation and prediction generation for the ARFIMA model, which were carried out with the use of the R statistical software.

The remainder of the paper is structured as follows. Section 2 briefly describes current financial asset forecasting approaches, including hybrid methods. Section 3 presents the data and illustrates applied data transformations. Section 4 contains a thorough description of research methodology, with special attention to the techniques of models' hybridization. Section 5 presents the derived results and performance of the created trading strategies. Section 6 summarizes the obtained findings and proposes a few extensions of the performed study. 

\section{Literature overview}

Efficient Market Hypothesis (Fama, 1970)  states that current prices of financial instruments fully reflect all available information.  EMH distinguishes three various forms of efficiency: weak, semi-strong, and strong, which differ in the applied sets of information. The weak form of the hypothesis takes into consideration the historical price or return series of the underlying asset. According to EMH in weak form, employing techniques focused on past price movements, like technical analysis or some predictive model, should not result in generating a trading strategy that would manage to systematically outperform the market. Verification of the EMH constitutes a vast body of financial literature, and the presented results are rather inconclusive. Especially in a case of less mature and more volatile markets, for example cryptocurrencies, the findings suggest a presence of periods of no efficiency (Khurseed et al. 2020; López-Martín et al., 2021). Moreover, the verification of EMH is also performed through the application of various types of predictive models to forecast future price values and create potentially profitable trading strategies. 

\subsection{Econometric models}

The classical approach to financial time series forecasting is represented by econometric models. Within this category, the ARIMA model (Box and Jenkins, 1976) presents an example of a frequently applied technique. Kobiela et al. (2022) applied ARIMA to forecast the values of the most liquid sectors of the NASDAQ stock exchange. Predictions were generated for different time horizons with the aim of simulating the behaviour of individuals with different investment preferences. In the case of long-term forecasting, the ARIMA model managed to outperform the LSTM network based on the price series as the only feature. Opposite results were reported for one-day ahead forecasting. 

However, in most studies, ARIMA serves as the benchmark model and point of reference for more advanced approaches. Vo and Ślepaczuk (2022) constructed a hybrid ARIMA-SGARCH model to create algorithmic investment strategies for the S\&P 500 index. The obtained results, both in terms of error metrics and trading indicators, demonstrated the superior performance of the extended architecture. However, it is worth noting that ARIMA managed to outperform the simple Buy\&Hold strategy. The S\&P 500 index was also used in the study by Kijewski and Ślepaczuk (2024). Investment strategies based on LSTM model predictions and signals generated by classical trading strategies generally yielded better results than the ARIMA model. Kumar (2010) compared the performance of the ARIMA and ANN models by generating short-term forecasts of the USD/JPN exchange rate. ANN proved to be a more accurate and robust approach in terms of the penalty-based and non-penalty-based evaluation criteria. Another field of research covers the prediction of cryptocurrency price movements. Akyildirim et al. (2023) performed a comparative analysis of multiple machine learning techniques to forecast Bitcoin futures contracts at different time horizons and frequencies. The majority of advanced approaches consistently outperformed the standard ARIMA model for most configurations. Similar results, specifically, the superiority of LSTM over ARIMA in forecasting Bitcoin, were reported by McNally et al. (2018).

The ARFIMA model (Granger and Joyeux, 1980; Hosking, 1981) represents another employed technique in financial time series forecasting. Bhardwaj and Swanson (2006) performed a thorough empirical investigation concerning the applicability of classical econometric techniques to model the returns of major worldwide stock indices. ARFIMA proved to provide more accurate approximations of the unknown data-generating processes in comparison to non-fractional alternatives. These findings serve as an argument for long-memory processes in financial applications. Analogous results have been observed in emerging capital markets. The advantage of the ARFIMA model over the simpler ARIMA model was supported by empirical investigations of the Greek stock market (Barkoulas et al., 2000) and the markets of the Middle East and North Africa countries (Assaf, 2006). Chaâbane (2014) proposed an application of the ARFIMA model and the feedforward neural network to forecast hourly spot prices in the electricity market. The nonlinear machine learning technique offered a more appropriate approach compared to the ARFIMA model, which was only capable of capturing linear patterns in the data.

\subsection{Machine Learning models}

Machine learning models, with their ability to detect nonlinear patterns and capture complex relationships between variables, offer an alternative to traditional statistical techniques in financial time series forecasting. Based on the systematic review of forecasting literature by Bustos and Pomares-Quimbaya (2020), the SVM model was among the most commonly applied techniques in recent years. Kim (2003) applied the SVM model to predict daily price movements of the Korean stock index using a set of 12 technical indicators as explanatory features. To address potential nonlinear properties in the data, the Gaussian radial basis function was employed as the kernel. SVM managed to achieve stable results, with the Hit Ratio statistic oscillating around 57\%. This value was significantly higher than that of the benchmark model, represented by the backpropagation neural network.  Similar results demonstrating the superiority of SVM over alternative models, including ARIMA and neural networks, were obtained for other stock indices such as the S\&P 500 (Cao and Tay, 2001), Nikkei 225 (Huang et al., 2005), and NASDAQ (Ince and Trafalis, 2008). Additionally, Shen et al. (2012), using datasets containing the main US stock indices, demonstrated the applicability of the SVM model in the context of trading. The remarkable trading performance of the SVM model across a broad set of national stock indices was also confirmed by Grudniewicz and Ślepaczuk (2023).

A less frequently explored approach in the literature for forecasting financial time series using machine learning techniques is the XGBoost model (Chen and Guestrin, 2011). Nobre and Neves (2019) proposed an advanced model architecture based on XGBoost to predict directional price movements for various types of financial assets. The extended data preprocessing module encompassed Principal Component Analysis and Discrete Wavelet Transform methods to reduce the dimensionality of data and denoise the features. Except for the S\&P 500 index, the developed system clearly outperformed the simple Buy\&Hold strategy. A comparative analysis of various machine learning models for predicting monthly gold prices using a set of macroeconomic variables as explanatory features was presented by Jabeur et al. (2024). The XGBoost model demonstrated the highest forecasting accuracy, indicating its suitability for commodity market prediction. Moreover, the utilization of Shapley additive explanations (SHAP) facilitated the model's interpretability. However, it is important to underline that some studies indicated that XGBoost might present tendencies to overfit the time series data during the training process, causing poor out-of-sample performance (Mills et al., 2024). Relatively weak effectiveness of XGBoost in modelling Nvidia's stock returns was also reported by Chlebus et al. (2021).

A novel machine learning approach to model sequential data is represented by the deep recurrent neural network - LSTM. Fischer and Krauss (2018) implemented the LSTM model to generate trading signals using the stocks constituting the S\&P 500 index. The obtained results indicate the superior performance of the LSTM-based strategy in comparison to Buy\&Hold and other machine learning techniques. An advanced model architecture using a broad set of major worldwide stock indices was proposed by Bao et al. (2017). The constructed predictive framework was based on three stages: wavelet transform to denoise the data, stacked autoencoders to extract features, and LSTM to generate predictions. The developed approach yielded significantly strong results, both in terms of accuracy and trading performance. It is worth mentioning that the efficiency of LSTM strongly depends on the proper construction of the model's architecture and the hyperparameter tuning procedure (Michańków et al., 2022).

There is also a growing body of research focused on applying the LSTM model to cryptocurrency markets. Viéitez et al. (2024) created a set of various predictive models to forecast the Ethereum price at different time horizons. The explanatory features included different sources of data, including price data of Ethereum and other assets, sentiment analysis information, and Google statistics. The authors concluded that the performance of the LSTM model was satisfactory in terms of both accuracy and investment returns. A set of 7 cryptocurrencies was used in the classification approach by Kwon et al. (2019). According to the values of the F1-score metric, LSTM managed to outperform the alternative gradient boosting model. The results supported the suitability of the LSTM model to forecast highly volatile assets. An empirical comparative analysis of multiple machine learning techniques for the most liquid cryptocurrencies was also conducted by Bouteska et al. (2024). However, the LSTM-based approach failed to deliver appropriate results.

\subsection{Hybrid approaches}

Hybrid methods, which integrate multiple independent models, are based on the assumption that combining forecasts results in improved accuracy. The study by Zhang (2003) constitutes a classic example in this area of research. The hybrid ARIMA-ANN model was applied to various time series datasets, including the GBP/USD exchange rate. The empirical investigation proved that the combined approach managed to outperform its individual components. Kumar and Thenmozhi (2014) compared several hybrid architectures based on the ARIMA model to forecast daily returns of the Indian stock index. In terms of predictive accuracy and trading performance, each hybrid model achieved superior results compared to its simple counterparts. Moreover, ARIMA-SVM was the best-performing model, surpassing ARIMA-ANN and ARIMA-RF. The superior performance of the ARIMA-SVM hybrid model in predicting various individual stock prices was also reported by Pai and Lin (2005). The ARIMA and XGBoost models were employed in a hybrid methodology focused on forecasting the energy supply security level in China (Li and Zhang, 2018). This task proved challenging for individual techniques due to its high dynamics and multiple influencing factors. However, the hybrid ARIMA-XGBoost model successfully addressed the aforementioned challenges, as demonstrated by the low values of the forecasting error metrics. Forecasting of financial data was also carried out using the hybrid ARFIMA-LSTM model. Bukhari et al. (2020) decided to model individual stock data using a set of macroeconomic variables as dependent features. The proposed ARFIMA-LSTM model effectively extracted the relevant relationships between the variables, leading to improved predictive accuracy compared to individual methods. An alternative hybrid architecture for the LSTM model was proposed by Kashif and Ślepaczuk (2025). For all three applied equity indices, LSTM-ARIMA outperformed its single constituents in terms of trading performance metrics.

However, some studies suggested that combining individual models does not necessarily lead to improved accuracy. Taskaya-Temizel and Casey (2005) applied the hybrid ARIMA-ANN model using various datasets. The obtained result indicated that, in most cases, the individual components outperformed the hybrid architecture. The authors concluded that the reduced performance of hybrid models might be influenced if the relationship between linear and nonlinear components in the data is not additive, an assumption underlying a majority of hybrid models. Analogical findings based on the cryptocurrency time series data were proved by Dudek et al. (2024). Multiple individual models and their combinations were tested, including ARFIMA, RF, SVM, and LSTM. It was stated that incorporating hybrid techniques did not lead to significant improvements. This observation could be attributed to the high level of noise and the absence of clear patterns in the data. Hibon and Evgeniou (2005) performed an extensive analysis concerning the potential advantages of combining forecasting approaches. The main finding of their study is that an individual model might outperform the hybridization. However, the advantage of model combination lies in the fact that selecting a random hybrid model is less risky in terms of forecasting performance than selecting a random individual method.

\section{Data}

\subsection{Data fetching and preprocessing}

The data used in the study was downloaded using the Python \textit{yfinance} library. The empirical investigation is based on two of the most liquid financial assets in the market: the stock market index S\&P 500 (USA) and the cryptocurrency Bitcoin. To ensure a reliable model training procedure and an effective generalization to unseen data, long historical price series were collected and analyzed. For S\&P 500, the data period covers daily observations between 1 January 2002 and 31 December 2023. In the case of Bitcoin, the dataset is made up of daily observations from 1 January 2015 to 31 December 2023. The collected price series consists of 5537 and 3286 observations, respectively.

As a first step, the downloaded data were cleaned and examined for missing or incorrect observations. After ensuring data validity, the logarithmic returns, representing the dependent variable in this study, were computed using the following formula (Vo and Ślepaczuk, 2022):

\begin{equation}
    r_t = \ln\left(\frac{P_t}{P_{t-1}}\right) = \ln\left(P_t\right) - \ln\left(P_{t-1}\right)
\end{equation}

\noindent where:
\begin{itemize}
  \item $r_t$ - logarithmic return at time $t$
  \item $P_t$ - closing price of the asset at time $t$.
\end{itemize}

\begin{figure}[H]
    \centering
    \caption{Logarithmic returns of S\&P 500 and Bitcoin prices}
    \includegraphics[width=1\linewidth]{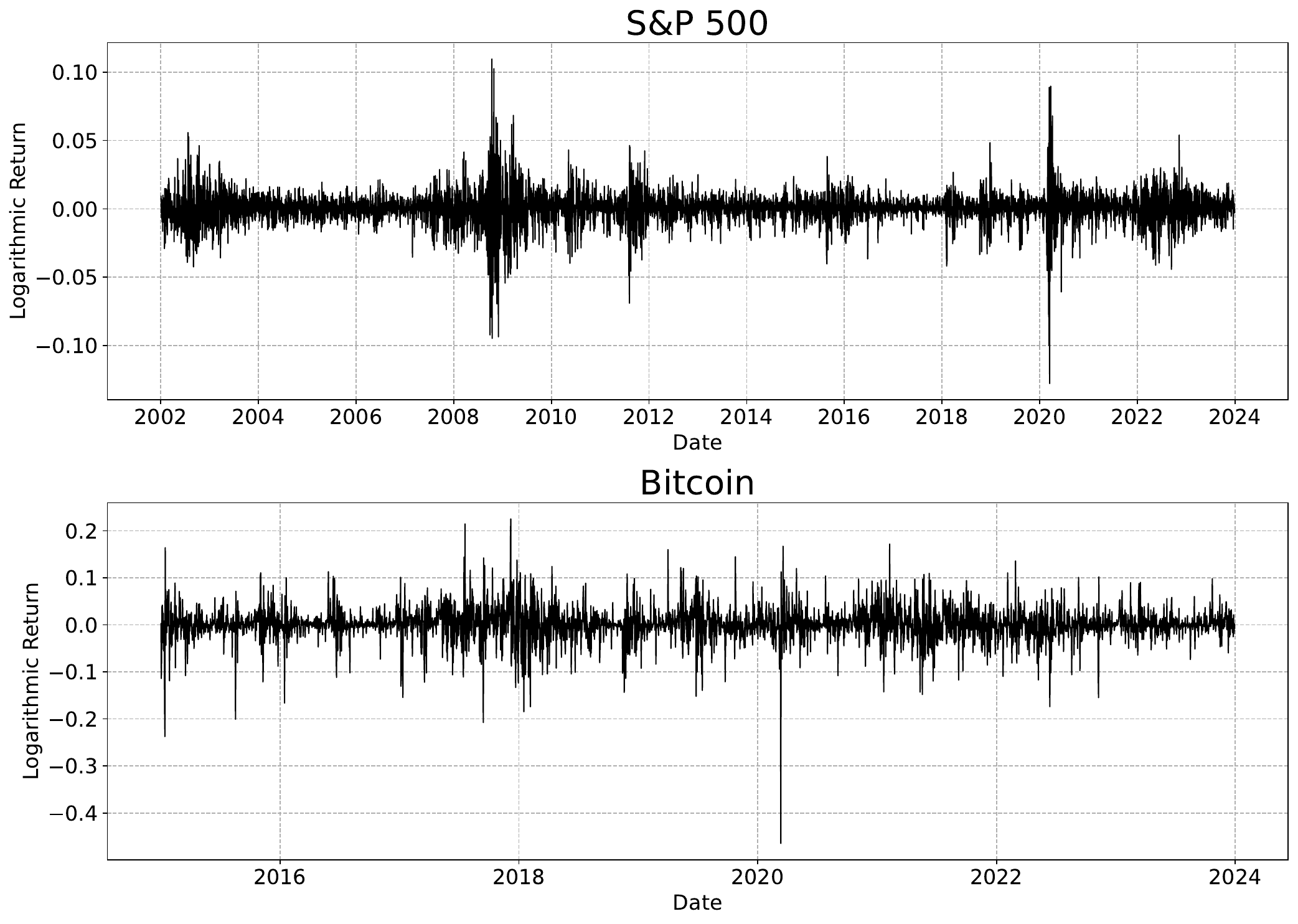}
    \vspace{1mm}
    \noindent
    {\raggedright\scriptsize
    \renewcommand{\baselinestretch}{0.9}\selectfont
    \textit{Note: S\&P 500 series covers the period between 1 January 2002 and 31 December 2023. For Bitcoin, the data covers the period from 1 January 2015 to 31 December 2023.}\par}

    \label{fig:log_ret}
\end{figure}

In the financial literature, there is rather a broad consensus regarding the use of logarithmic returns instead of simple returns. One of their main advantages is that they can be interpreted as continuously compounded returns, which facilitates their additivity over time (Hudson and Gregoriou, 2015). Figure \ref{fig:log_ret} presents the graphical representation of the levels and dynamics of the logarithmic return series for S\&P 500 and Bitcoin over the study period. It is visible that Bitcoin is characterized by significantly higher volatility, with greater amplitude of daily price movements. This observation aligns with the view of cryptocurrency markets as highly risky and speculative (Liu and Serletis, 2019). Moreover, the periods of increased volatility in the markets correspond to the 2008 global financial crisis and the outbreak of the COVID-19 pandemic in 2020.

\subsection{Descriptive statistics}

Table \ref{tab:desc_stat} presents the calculated descriptive statistics values for the logarithmic returns of S\&P 500 and Bitcoin for the whole study period. These results support the finding regarding increased Bitcoin volatility in comparison to S\&P 500. The values of minimum, first quartile, third quartile, and maximum statistics indicate that the distribution of Bitcoin returns is highly dispersed and asymmetric. Moreover, a higher value of standard deviation in comparison to S\&P 500 returns validates the evidence of Bitcoin as a highly volatile asset. The negative values of the skewness statistic indicate that the returns of both assets are left-skewed. This implies that investors are exposed to the risk of substantial daily losses, which are not adequately compensated by the corresponding gains. Additionally, the highly positive values of kurtosis suggest the leptokurtic distribution of the returns. This implies that, compared to the normal distribution, the tails are heavier, which increases the likelihood of extreme events. In a financial context, such a property reflects the higher probability of both substantial gains and losses.

\begin{table}[htbp]
\centering
\caption{Descriptive statistics for logarithmic returns of S\&P 500 and Bitcoin}
\label{tab:desc_stat}
\begin{tabular}{p{4cm} p{2cm} p{2cm}}
\toprule
\textbf{Descriptive statistic} & \textbf{S\&P 500} & \textbf{Bitcoin} \\
\midrule
Count & 5536 & 3285 \\
Min & -12.77\% & -46.47\% \\
1st quartile & -0.45\% & -1.22\% \\
Median & 0.07\% & 0.14\% \\
Arithmetic mean & 0.03\% & 0.15\% \\
3rd quartile & 0.58\% & 1.69\% \\
Max & 10.96\% & 22.51\% \\
Standard deviation & 1.22\% & 3.74\% \\
Skewness & -0.42 & -0.79 \\
Kurtosis & 11.49 & 11.60 \\
\bottomrule
\end{tabular}
\vspace{1mm}

\noindent
{\raggedright\scriptsize\textit{Note: S\&P 500 series covers the period between 1 January 2002 and 31 December 2023. For Bitcoin, the data covers the period from 1 January 2015 to 31 December 2023.}\par}
\end{table}

\subsection{Data sampling}

\begin{figure}[H]
    \centering
    \caption{Cross-validation scheme for machine learning models (S\&P 500)}
    \includegraphics[width=1\linewidth]{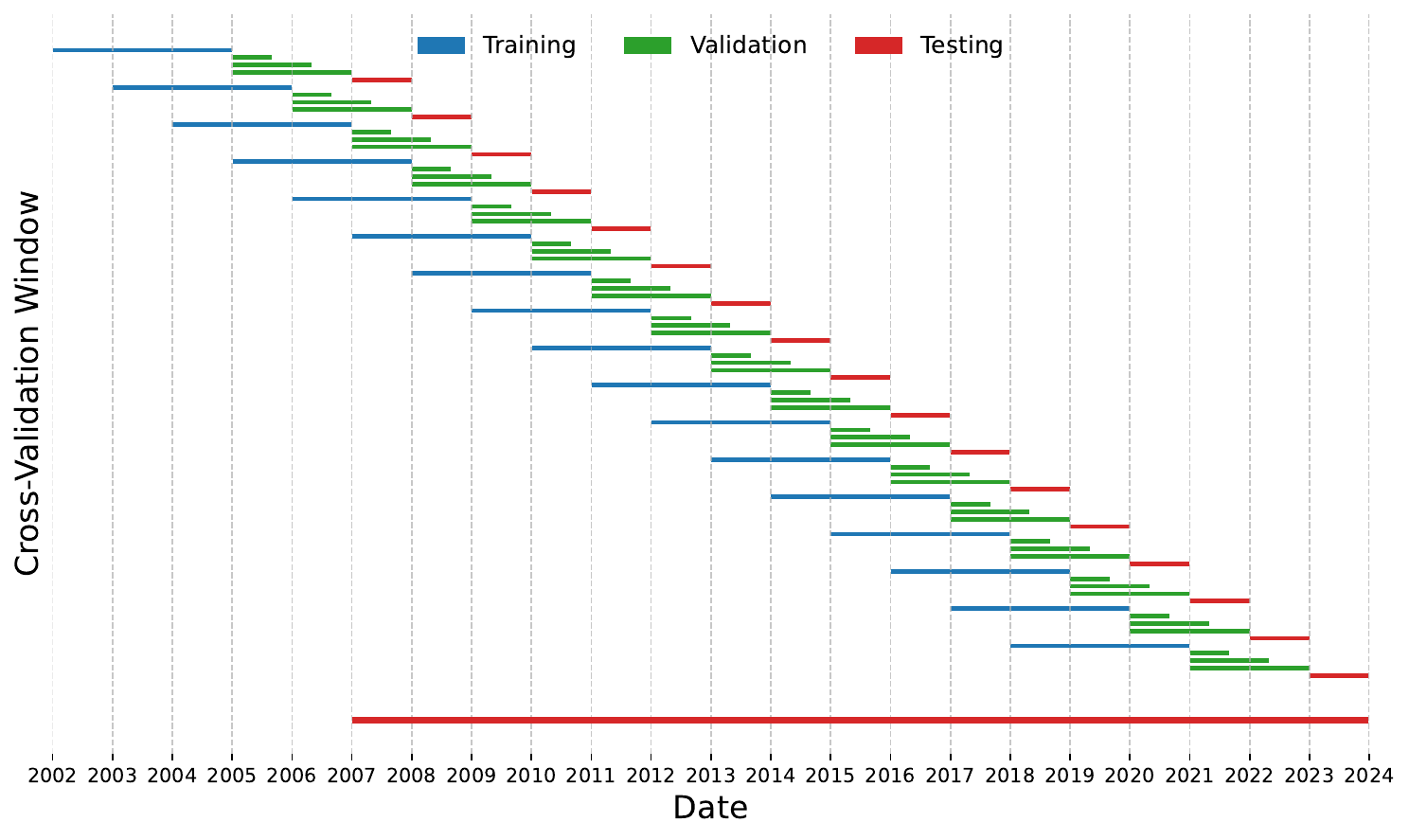}
    \vspace{1mm}
    \noindent
    {\raggedright\scriptsize
    \renewcommand{\baselinestretch}{0.9}\selectfont
    \textit{Note: Figure presents the cross-validation scheme applied to training machine learning models for the S\&P 500 index. Each window consists of a three-year training period, three validation periods of 8, 16, and 24 months, and a one-year testing period. The red line at the bottom indicates the total length of the testing period spanning between 1 January 2007 and 31 December 2023.}\par}
    \label{fig:cv}
\end{figure}

Due to the temporal dependence structure present in time series data, common cross-validation techniques to train machine learning models are not applicable. To prevent data leaking, the time order of the observations should be preserved. For this reason, this study employs cross-validation on a rolling basis with a dynamically adjusted window. The technique allows for sequential recalibration of the predictive model using only the latest data. This property allows the model to adapt to changing market conditions and recognize new patterns. However, this study proposes a novel time series cross-validation scheme that constitutes a variation of the approach introduced by Choi et al. (2024). Figure \ref{fig:cv} presents the construction of the three-fold cross-validation technique for the S\&P 500 stock index. Each window covers the six-year period with the separated training, validation, and testing sets. The length of each segment corresponds to the three-year, two-year, and one-year periods, respectively. However, the validation set is additionally divided into three separate sections of 8, 16, and 24 months. For each iteration, the window moves one year forward. The optimal set of hyperparameters is selected based on the average performance across three distinct validation folds. The objective of the proposed three-fold cross-validation method is to enhance the robustness of the hyperparameter selection process.

A similar approach is applied to the Bitcoin time series, although the window size is modified. Each window consists of a two-year training period, three validation periods of 4, 8, and 12 months, and a six-month testing period. In each iteration, the window shifts forward by six months. The total length of the testing period for Bitcoin spans between 1 January 2018 and 31 December 2023.

\section{Research methodology}

\subsection{Econometric models}

\subsubsection{Autoregressive Integrated Moving Average model - ARIMA}

Autoregressive Integrated Moving Average model (ARIMA), introduced by Box and Jenkins (1976), constitutes one of the most basic and commonly used statistical models in time series analysis. It is based on the assumption that the relationship between observations in the series is described by their autocorrelation, meaning that their past values and past prediction errors may influence current observations (Kumar, 2010). The ARIMA model is constructed out of three separate parts: Autoregressive model (AR), Integration (I), and Moving Average model (MA). In order to apply ARIMA methodology, the underlying series is required to be either stationary or to be first transformed into the stationary form (Kobiela et al., 2022). The stationary structure of the time series is characterized by the stability of its mean, variance, and covariance over time (Granger and Newbold, 1974). This process is carried out in the Integration (I) component of the model through the differencing operation. Differentiation is performed with the use of the lag operator $B$ (Harvey, 1990):

\begin{align}
    BX_{t} &= X_{t-1} \\
    B^{d} X_{t} &= X_{t-d}
\end{align}

\noindent where:
\begin{itemize}
  \item $X_i$ – value of the series $X$ observed at time $i$
  \item $d$ – order of lag operator.
\end{itemize}

The Autoregressive (AR) model captures a linear dependence structure between the current, previous values and a random error component. The AR(p) process can be denoted in the following way: 

\begin{equation}
    X_t = \phi_1 X_{t-1} + \phi_2 X_{t-2} + ... + \phi_p X_{t-p} + \epsilon_t
\end{equation}

\noindent where:
\begin{itemize}
  \item $\phi_1, ..., \phi_p$ – parameters of the autoregressive model of order p
  \item $\epsilon_i$ – random error component (white noise) at time $i$.
\end{itemize}

Similarly, the Moving Average process (MA) utilizes the lagged white noise error terms to describe the current value of the time series (Vo and Ślepaczuk, 2022). The MA(q) model may be formulated as:

\begin{equation}
    X_t = \mu + \epsilon_t + \theta_1 \epsilon_{t-1} + \theta_2 \epsilon_{t-2}  +...+ \theta_q \epsilon_{t-q} 
\end{equation}

\noindent where:
\begin{itemize}
  \item $\theta_1, ..., \theta_q$ – parameters of the moving average model of order q
  \item $\mu$ – mean value of the series.
\end{itemize}

By combining the presented components, the ARIMA(p,d,q) model is constructed, where p, d, and q denote the orders of the autoregressive, integration, and moving average parts, respectively. It can be written as:

\begin{equation}
    (1-\sum_{i=1}^{p} \phi_i B^i)(1-B)^d X_t = (1+\sum_{i=1}^{q} \theta_i B^i)\epsilon_i
\end{equation}

In the classical approach, the process of finding the optimal orders of the ARIMA(p,d,q) model was conducted based on the Box-Jenkins approach. The methodology was divided into three stages: model identification, parameter estimation, and diagnostic checking. Candidate values for model parameters were selected after analyzing the sample autocorrelation and partial autocorrelation functions (Newbold, 1975). However, this approach was characterized by some significant drawbacks. The model specification process was primarily based on the researcher's judgment rather than on an automated mathematical framework (De Gooijer and Hyndman, 2006). Nowadays, the estimation of the ARIMA model is often based on the use of information criterion estimators, which assess the goodness of fit of the applied model while penalizing overfitting (Stoica and Selen, 2004). In this study, the Akaike Information Criterion (AIC) was chosen as the applied method for the model selection process. For forecasting purposes, the most appropriate model is selected using the in-sample part of the data. Then, the estimated values of the parameters are employed to generate predictions for the unseen out-of-sample observations.

\subsubsection{Autoregressive Fractionally Integrated Moving Average model - ARFIMA}

The estimated autocorrelation function between the observations in the series often exhibits patterns that raise questions about whether the ARIMA model should be utilized as the appropriate data-generating process (Granger and Ding, 1996). Specifically, the ARIMA process is characterized by an exponential decay structure of the autocorrelations. On the other hand, many empirical datasets exhibit a hyperbolic decay structure, which drops significantly slower (Koustas and Serletis, 2005). This property is described as the long-term memory of the time series. Formally, the series reveals long memory characteristics if some innovation $\epsilon_t$ at time $t$ continues to exert influence on $X_{t+k}$ for larger $k$, relative to the behavior predicted by the ARIMA model (Granger and Ding, 1996). 

To address the described challenges, Granger and Joyeux (1980) and Hosking (1981) introduced a novel time series model: the Autoregressive Fractionally Integrated Moving Average model (ARFIMA). The ARFIMA model constitutes an extension of the ARIMA methodology by incorporating fractional differencing. Fractional differentiation allows for the integration of time series for non-integer orders of parameter $d$. Consequently, the fractional integration operator (FI), using the binomial theorem, may be formulated as (De Prado, 2018):

\begin{equation}
\begin{aligned}
\left( 1 - B \right)^d X_t &= X_t - d X_{t-1} + \frac{d(d-1)}{2!} X_{t-2}+ \cdots \\
&+ (-1)^k \prod_{i=0}^{k-1} \frac{(d-i)}{k!} X_{t-k} + \cdots
\end{aligned}
\end{equation}

The ARFIMA(p,d,q) model, with $d$ taking any real non-integer value and p and q denoting the orders of the autoregressive and moving average components, can be stated as: 

\begin{equation}
    (1-\sum_{i=1}^{p} \phi_i B^i)(1-B)^d X_t = (1+\sum_{i=1}^{q} \theta_i B^i)\epsilon_i
\end{equation}

For the estimated value of parameter $d \in (-0.5, 0.5)$, the series $X$ exhibits stationary and invertible properties. In particular, for $d \in (0, 0.5)$, the ARFIMA(p,d,q) process is characterized by the long-term behaviour (Hosking, 1981). Since many financial time series reveal dependencies for larger time horizons, the ARFIMA model demonstrates potential advantages in comparison to the short-term memory ARIMA model (Bhardwaj and Swanson, 2006). Many researchers undertake the topic of the empirical investigation of the presence of long-term memory in stock return series. This task is usually performed with a set of statistical tests focused on detecting the underlying dependence structure in the analyzed series. While some papers seem to suggest the presence of long memory in the stock returns (Assaf, 2006; Floros et al., 2007), other studies find little evidence in support of this (Lo, 1991; Cheung and Lai, 1995). In the case of Bitcoin, the persistence of long memory has been shown to be associated with the periods of increased market inefficiency (Jiang et al., 2018; Wu et al., 2022).

\subsection{Machine Learning and Deep Learning Models}

\subsubsection{Support Vector Machines - SVM}

Support Vector Machines (SVM) constitutes a supervised machine learning model applied to classification and regression tasks and motivated by statistical learning theory (Bennett and Campbell, 2000). SVM is characterized by high generalization properties due to the structural minimization principle, which prevents the model from overfitting. Moreover, the obtained solution, derived by solving a constrained quadratic programming task, is proved to always be globally optimal (Huang et al., 2005). This property ensures the stability and reproducibility of the results. The first form of the SVM model, introduced by Cortes and Vapnik (1995), was applied for a two-group classification task where the input vectors are non-linearly transformed to a high-dimensional feature space. The goal of the algorithm is to identify an optimal linear hyperplane separating the two classes in that feature space. The mapping of inputs into the higher dimensions is performed with so-called kernel functions (Kim, 2003). In subsequent years, SVM has been further developed, which enabled employing the model in novel applications, including non-separable data, multi-class classification problems, and regression tasks (Bennett and Campbell, 2000).

The regression alternative of SVM is represented by the Support Vector Regression (SVR) model. SVR employs the $\epsilon$-insensitive loss function, intending to maintain the fitting error below a specified threshold (Cocco et al., 2021). Consequently, the goal of the SVR method is to approximate the underlying function in the form presented in Equation \ref{eq:approx}. It is achieved by minimizing the objective function in Equation \ref{eq:obj}, also called the primal formula, subjected to the constraints (MathWorks, 2025). 

\begin{equation}
f(X) = X' \beta + b
\label{eq:approx}
\end{equation}

\begin{equation}
J(\beta) = \frac{1}{2} \beta' \beta + C  \sum_{i=1}^N (\xi_i + \xi_i^*)
\label{eq:obj}
\end{equation}

\text{subject to:}
\begin{align*}
\forall i: \, y_i - (X_i' \beta + b) &\leq \epsilon + \xi_i \\
\forall i: \, (X_i' \beta + b) - y_i &\leq \epsilon + \xi_i^* \\
\forall i: \, \xi_i, \xi_i^* &\geq 0
\end{align*}

\noindent where:
\begin{itemize}
  \item $X$ – matrix of features
  \item $\beta$ – vector of coefficients
  \item $b$ – intercept coefficient
  \item $C$ – positive constant controlling the magnitude of penalty imposed on observations placed outside the $\epsilon$ margin
  \item $\xi_i, \xi_i^*$ – slack variables describing the magnitude of the acceptable regression errors
  \item $\epsilon$ – margin of no error
  \item $y$ – vector of targets.
\end{itemize}

As mentioned earlier, the family of SVM models employs kernel functions to map features into a higher-dimensional space. The most commonly used, also utilized in this study, include the linear, polynomial, and radial basis function kernels (Cocco et al., 2021). The latter two allow for modeling nonlinear relationships in the data.

\subsubsection{eXtreme Gradient Boosting - XGBoost}

In the machine learning domain, boosting constitutes an ensemble method employing a number of weak learners, that is, models that are only slightly more accurate than random guessing, to construct a single strong learner (Schapire, 1999). Boosting is based on the idea of iteratively fitting subsequent weak learners to the readjusted dataset. At each step, more weight is given to the observations characterized by the greatest prediction error values. At the end of the training process, outputs of the weak learners are integrated in a weighted manner. The final output is constructed using either voting or averaging for the classification and regression tasks, respectively (Nobre and Neves, 2019). One of the earliest successful implementations of boosting was represented by the Adaboost algorithm, utilizing the decision stumps as the weak learners (Freund and Schapire, 1997). 

The eXtreme Gradient Boosting model (XGBoost) represents an advanced, scalable, and computationally efficient technique implemented by Chen and Guestrin (2011). XGBoost constitutes an optimized implementation of the gradient boosting algorithm, supporting the utilization of various objective functions (Chatzis et al., 2018). In addition, XGBoost is also equipped with various regularization techniques to prevent overfitting. Ensembling is carried out using the Classification and Regression Tree model (CART) as the base learners. Simplified mathematical representation of the XGBoost mechanism may be stated in the following way (Chen and Guestrin, 2011):

\begin{equation}
 \hat{y}_i = \sum_{k=1}^{K} f_k(X_i)
\label{eq:xgb1}
\end{equation}

\begin{equation}
\mathcal{L}^{(t)} = \sum_{i=1}^{n} l\left( y_i, \hat{y}_i^{(t-1)} + f_t(X_i) \right) + \Omega(f_t)
\label{eq:xgb2}
\end{equation}

\noindent where:
\begin{itemize}
  \item $K$ – number of trees
  \item $f_k$ – each independent tree structure
  \item $\hat{y}_i$ – output prediction for observation $i$
  \item $\hat{y}_i^{(t-1)}$ – output prediction for observation $i$ at $t-1$ iteration
  \item $l$ – training loss function
  \item $\Omega$ – regularization term.
\end{itemize}

Equation \ref{eq:xgb1} presents the process of generating model predictions by adding independent weak learners. Equation \ref{eq:xgb2} contains the objective function, which incorporates the regularization term. The training process is conducted by minimizing the objective function through iteratively adding subsequent learners. 

\subsubsection{Long Short-Term Memory - LSTM}

The Long Short-Term Memory model (LSTM) was originally proposed by Hochreiter and Schmidhuber (1997), and its architecture has been further extended in subsequent years (Gers et al., 2000; Gers et al., 2002; Greff et al., 2016). The LSTM model represents a category of recurrent neural networks (RNN) designed to process sequential data, including time series. Consequently, RNNs found applications in many areas of research, including natural language processing, speech recognition, energy load forecasting, and stock market prediction (Al-Selwi et al., 2024). However, basic RNN models appeared to be ineffective in learning long-term dependencies between the lagged observations. This feature is referred to as the vanishing gradient problem and is caused by the insufficient weight changes during the training process using the backpropagation algorithm (Hochreiter, 1998). 

\begin{figure}[H]
\centering
\caption{Internal flow of information inside the LSTM network memory cell}
\includegraphics[width=0.6\textwidth]{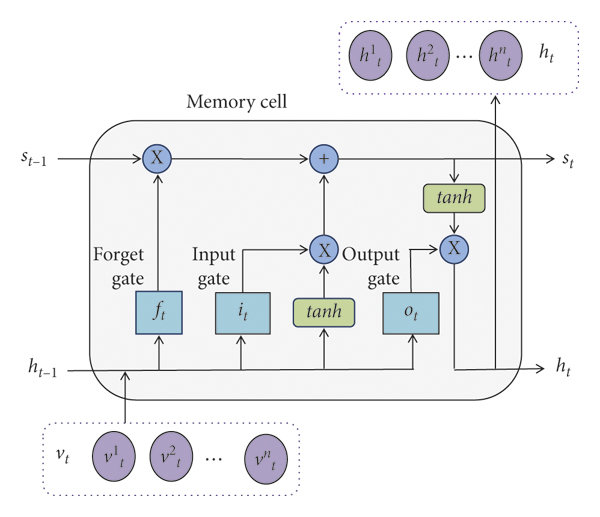}
\vspace{1mm}

\noindent
{\raggedright\scriptsize\textit{Source: Yang et al. (2020), \url{https://www.researchgate.net/figure/The-structure-of-LSTM-memory-cell_fig5_342998863}}\par}

\label{fig:lstm_cell}
\end{figure}

To overcome the shortcomings of simple recurrent neural networks, the LSTM model possesses a more complex architecture. Similar to basic feedforward networks, LSTM is composed of three types of layers: the input layer, one or more hidden layers, and the output layer. Unlike other networks that made use of neurons, the hidden layers of LSTM are constructed out of parallel stacks of memory cells (Bao et al., 2017). The architecture of the memory cell is based on the input gate - $i_t$, the forget gate - $f_t$, and the output gate - $o_t$. They control the flow of knowledge inside the memory cell through appending new information, resetting the memory, and returning the output (Gers et al., 2000). The mathematical operations are performed using the sigmoid and hyperbolic tangent activation functions. Figure \ref{fig:lstm_cell} presents the architecture and the internal mechanism of the LSTM network memory cell.

The computed operations inside the memory cell may be mathematically described by the following set of equations (Fischer and Krauss, 2018; Bao et al., 2017):

\begin{equation}
\begin{aligned}
f_t &= \sigma(W_{fv} v_t + W_{fh} h_{t-1} + b_f) \\
\tilde{s}_t &= \tanh(W_{{\tilde{s}}v} v_t + W_{{\tilde{s}}h} h_{t-1} + b_{\tilde{s}}) \\
i_t &= \sigma(W_{iv} v_t + W_{ih} h_{t-1} + b_i) \\
s_t &= f_t \odot s_{t-1} + i_t \odot \tilde{s}_t \\
o_t &= \sigma(W_{ov} v_t + W_{oh} h_{t-1} + b_o) \\
h_t &= o_t \odot \tanh(s_t)
\end{aligned}
\end{equation}

At the beginning, the activation value of the forget gate $f_t$ is calculated using the output of the memory cell at the previous timestamp $h_{t-1}$ and current input vector $v_t$. At this stage, a decision is made regarding which information should be removed. The next equations present the computation of the candidate value for the updated memory cell ${\tilde{s}}_t$ and the activation value of the input gate $i_t$. This phase determines which new information will update the cell state. By combining the previously calculated values, the new cell state $s_t$ is obtained. In the end, the activation value of the output gate $o_t$ is computed, and the output of the memory cell at the current timestamp $h_t$ is passed on. Matrices $W$ and bias vectors $b$ are updated during the training process using the Backpropagation Through Time (BPTT) algorithm (Greff et al., 2016). Owing to the described mechanism, the LSTM model presents the ability to effectively handle long-term dependencies, find subtle patterns, and forecast unseen values of the time series process with enhanced accuracy (Viéitez et al., 2024).

\subsection{Hybridization methodology}

Due to the low signal-to-noise ratio and substantial uncertainty inherent in financial data, the task of financial time series forecasting remains a significant challenge (De Prado, 2015). Moreover, data data-generating process might exhibit both linear and nonlinear dependence structures that additionally may switch over time (Terui and Van Dijk, 2002). For this reason, the linear econometric time series models, as ARIMA or ARFIMA, might be inadequate to approximate complex functions containing a nonlinear component (Li and Zhang, 2018). This property led to the development of novel forecasting techniques based on combining multiple models with different characteristics. The hybrid methodology is based on the assumption that combining diverse individual models allows for obtaining more accurate predictions. Especially, the merge of the linear and nonlinear models provides a solid theoretical background supporting the potential effectiveness of this operation, as each model is responsible for extracting different information from data. Consequently, the hybridization of econometric models with nonlinear machine learning models has been studied in the literature (Pai and Lin, 2005). In this study, each econometric model (ARIMA and ARFIMA) is combined with an individual machine learning model (SVM, XGBoost, and LSTM). Moreover, two distinct methodologies of creating hybrid models are compared.

Following the hybridization method introduced by Zhang (2003), the time series data-generating process may be described as the combination of the linear and nonlinear components

\begin{equation}
y_t = L_t + N_t,
\end{equation}

\noindent where $L_t, N_t$ denote the linear and nonlinear parts, respectively. The initial step of forecasting is performed by applying the linear statistical model. Its responsibility is to extract the linear dependence structure. As a result, the residual $e_t$ is obtained through the following operation

\begin{equation}
e_t = y_t - \hat{L}_t,
\end{equation}

\noindent where $\hat{L}_t$ represents the prediction of linear model. If the econometric model managed to properly extract the linear component from the data, the residual $e_t$ ought to contain only the nonlinear segment and the error term. In the next phase, the nonlinear machine learning model is used to model the residuals. With $n$ lagged residuals employed as the model features, 

\begin{equation}
\hat{N}_t = f(e_{t-1}, e_{t-2},...,e_{t-n}),
\end{equation}

\noindent where $\hat{N}_t$ denotes the prediction of nonlinear model and $f$ constitutes the nonlinear function applied by the model. In the end, the predictions of both models are combined, and the final forecast $\hat{y}_t$ of the time series at time $t$ is generated

\begin{equation}
\hat{y}_t = \hat{L}_t + \hat{N}_t.
\end{equation}

The second applied approach of hybridization might be described by the following formula: 

\begin{equation}
\hat{y}_t = f(y_{t-1}, y_{t-2},...,y_{t-n}, \hat{L}_t).
\end{equation}

\noindent So, the final prediction at the moment $t$ is derived by inputting $n$ lagged observations of the time series process and the forecast of the linear model $\hat{L}_t$ for day $t$ to the nonlinear machine learning model. The predictive model makes use of the original observations and is additionally extended by the linear component extracted using econometric method. The potential advantage of this method lies in the fact that it does not assume the additive relation between the linear and nonlinear components. This hybridization technique has been utilized, among others, by Kashif and Ślepaczuk (2025).

\subsection{Error metrics}

Two different error metrics are employed to thoroughly evaluate the constructed models' forecasting performance. This division allows for a deeper insight into the reliability of the computed predictions. The selected metrics are presented in the following order:

Root Mean Squared Error (RMSE)

\begin{equation}
\text{RMSE} = \sqrt{ \frac{1}{N} \sum_{i=1}^{N} \left( \hat{y}_i - y_i \right)^2 }
\end{equation}

Mean Absolute Error (MAE)

\begin{equation}
\text{MAE} = \frac{1}{N} \sum_{i=1}^{N} \left| \hat{y}_i - y_i \right|
\end{equation}

\noindent where:
\begin{itemize}
  \item $y_i$ – actual value for observation $i$
  \item $\hat{y}_i$ – predicted value for observation $i$
  \item $N$ – number of observations.
\end{itemize}

\subsection{Trading strategies construction}

In order to empirically investigate the practical application of the applied models, the generated predictions are used as the basis for trading strategies. The financial literature refers to this procedure as \textit{backtesting}. The backtest may be defined as a historical simulation of the performance of the algorithmic investment strategy, that is, its potential gains and losses, applied to the out-of-sample part of the data (De Prado, 2018).  Similarly to the area of machine learning, the process of backtesting is exposed to the risk of overfitting, where the applied strategy captures a specific case instead of the general structure of the market (Bailey et al., 2014). Backtest overfitting constitutes the main reason why many models that yield promising results on paper fail when applied to the real market (Bailey et al., 2016).

To minimize the risk of backtest overfitting, the employed transformation of predictions into trading signals is relatively simple and presents itself in the following way:

\begin{equation}
\text{Signal} =
\begin{cases}
1 & \text{if } \hat{y}_{i+1} > c \\
0 & \text{if } \left| \hat{y}_i\right| \leq c \\
-1 & \text{if } \hat{y}_{i+1} < - c
\end{cases}
\end{equation}

\noindent where:
\begin{itemize}
  \item $\hat{y}_{i+1}$ – the predicted value of return for observation $i+1$ made at the end of the day $i$
  \item $c$ – transaction costs (formulated in percentages)
\end{itemize}

Based on the generated signals, the appropriate market positions are determined. Two relatively basic trading strategies are employed in this paper: \textit{\textit{Long-Short}} and \textit{\textit{Long Only}}. To realistically reflect the mechanisms of financial markets, each trading operation is associated with adequate transaction costs. The process of opening a new position or modifying an existing one is performed only if potential profits from the trade surpass the transaction costs. The \textit{\textit{Long-Short}} strategy allows for holding either a long or a short position in an asset. To generate a buy signal, the predicted value of the next day's return must exceed the transaction costs. Similarly, in the case of opening a short position. Unless the expected return surpasses the transaction costs, no operation is performed, and the previous position is maintained. Changing the long position into short, or the other way around, is allowed. However, the operation is associated with double transaction costs because it requires closing the previous position and opening the new one. In the \textit{\textit{Long Only}} framework, there is no possibility of holding a short position. So, a sell signal is executed only if the position has already been opened. Otherwise, no action is taken.

\subsection{Trading performance metrics}

Performance evaluation of different constructed models is conducted using a set of various trading efficiency metrics. The indicators underline various aspects of the applied strategies, including their returns, risk, and risk-adjusted returns. Based on Ryś and Ślepaczuk (2019) and Gómez and Ślepaczuk (2021), the following section presents and briefly describes the metrics employed in the study.

Annualized rate of return (ARC) - the simplest and most frequently used meitric in the algorithmic investing literature. It reflects the average annual return on investment over multiple years, accounting for compounding of returns. The metric provides no information about the risk associated with the applied strategy.
\begin{equation}
    \text{ARC} = \left( \frac{V_{t_2}}{V_{t_1}} \right)^{\frac{1}{D(t_1, t_2)}} - 1
\end{equation}

\noindent where:
\begin{itemize}
    \item \( V_{t_i} \) – value of the asset at time \( t_i \)
    \item \( D(t_1, t_2) \) – number of years between \( t_1 \) and \( t_2 \)
\end{itemize}

Annualized standard deviation (ASD) - the empirical standard deviation of returns generated by the strategy. An indicator of the investment strategy's volatility. The number of trading days used in the presented formula depends on the class of the financial asset. In the study, the values of 252 and 365 days are applied to the S\&P 500 index and Bitcoin, respectively (Bieganowski and Ślepaczuk, 2024). 
\begin{equation}
    \text{ASD} = \sqrt{T} \cdot \sqrt{\frac{1}{N-1} \sum_{i=1}^N (r_i - \bar{r})^2}
\end{equation}

\noindent where:
\begin{itemize}
    \item \( T \) – characteristic for the individual asset number of trading days in a year
    \item \( N \) – total number of trading days
    \item \( r_i \) – assets' percentage return on day \( i \)
    \item \( \bar{r} \) – assets' average daily percentage return
\end{itemize}

Maximum drawdown (MD) - the maximum percentage decline in the portfolio value during the investment duration. A stable investing strategy is characterized by the small value of the metric, which indicates its resistance to the volatile market conditions (Grudniewicz and Ślepaczuk, 2023). Since investors are typically characterized by risk aversion, the maximum drawdown indicator constitutes a suitable tool to assess the potential severity of losses (Geboers et al., 2023).
\begin{equation}
    \text{MD} = \sup_{x, y \in [t_1, t_2]^2 : x \leq y} \frac{P_x - P_y}{P_x}
\end{equation}

\noindent where:
\begin{itemize}
    \item \( P_t \) – equity line level at time \( t \)
\end{itemize}

Information ratio (IR) - metric calculated as the ratio of the annualized rate of return to the annualized standard deviation (Sharpe, 1966). The metric represents a more holistic approach to the performance evaluation as it considers two components of the trading strategy: achieved returns and associated risk. High values indicate that the strategy generates substantial returns with relatively low volatility. The IR metric facilitates the comparison of the effectiveness of various investment strategies both with each other and against a benchmark.

\begin{equation}
    \text{IR} = \frac{\text{ARC}}{\text{ASD}}
\end{equation}

Adjusted information ratio ($\text{IR}^{*}$) - the extended version of the Information ratio metric. It takes into account the additional risk factor of the implemented trading strategy: maximum drawdown. The metric provides an alternative yet complementary approach to computing the risk-weighted gains of the investment (Kosc et al., 2019).
\begin{equation}
    \text{IR}^{*} = \frac{\text{ARC}^{2} \cdot \text{sign}(\text{ARC})}{\text{ASD} * \text{MD}}
\end{equation}

Sortino ratio (SR) - another kind of risk-adjusted return metric. It substitutes the annualized standard deviation with the annualized standard deviation of downside movements as a measure of risk. Unlike the IR metric, the Sortino ratio excludes positive returns when calculating volatility. This modification reflects the preferences of the investors, which distinguish between upside and downside volatility of their investment strategy (Kolbadi and Ahmadinia, 2011).

\begin{equation}
    \text{SR} = \frac{\text{ARC}}{\text{ASD}^{-}}
\end{equation}

\noindent where:
\begin{itemize}
    \item \( \text{ASD}^{-} \) – annualized standard deviation of downside deviations (negative returns)
\end{itemize}

\section{Empirical results}

\subsection{Investment strategies for the S\&P 500 index}

\begin{figure}[H]
\centering
\caption{Equity lines for S\&P 500 and \textit{Long-Short} trading strategy}
\includegraphics[width=1\textwidth]{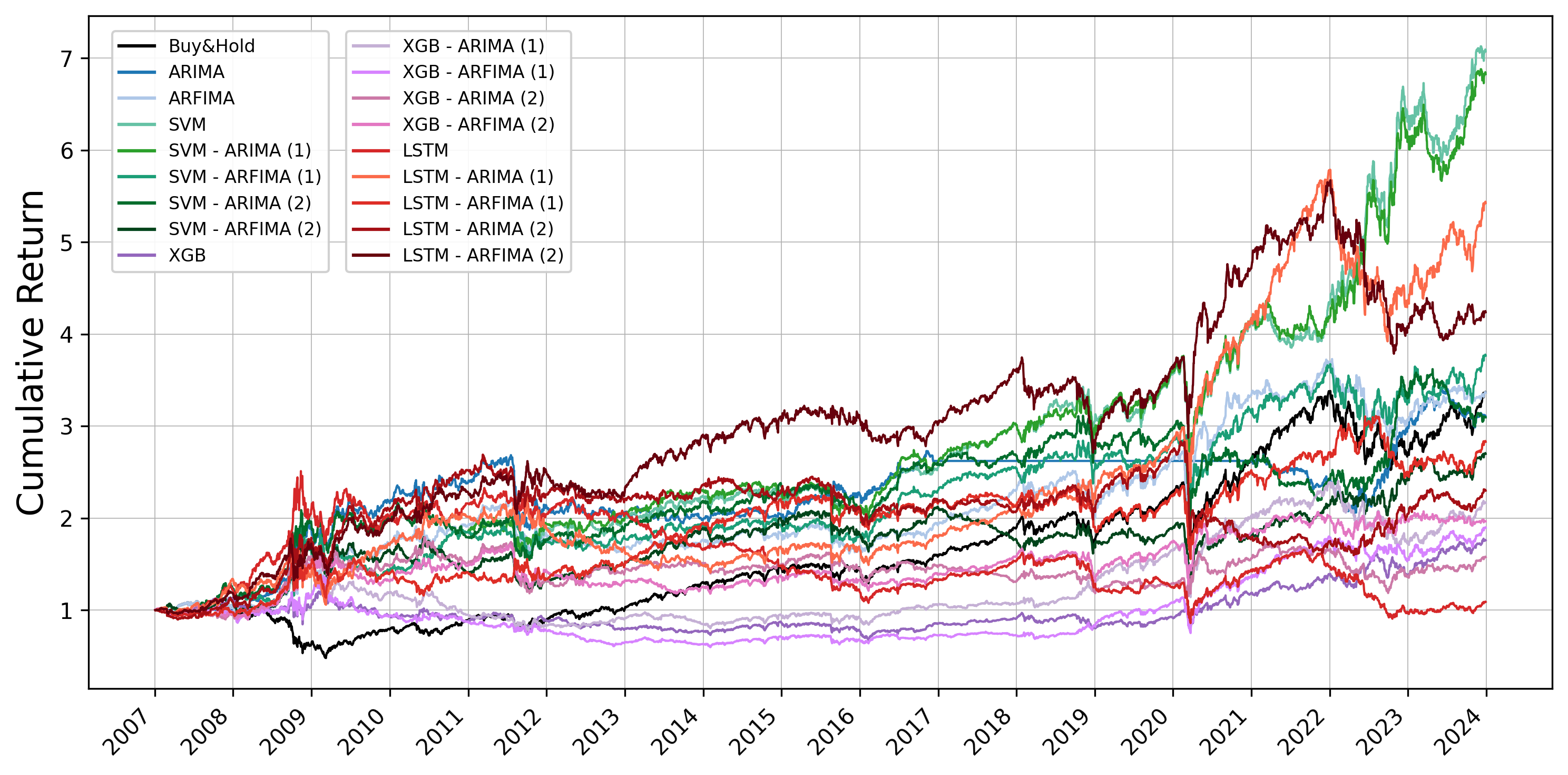}
\caption*{\begin{minipage}[t]{\textwidth}
\raggedright \scriptsize\textit{Note: All equity lines refer to predictive models out-of-sample forecasts. The first line represents the benchmark Buy\&Hold strategy. Annotation (1) denotes the hybridization technique of inputting the prediction of the econometric model to the machine learning model, while (2) indicates the hybrid methodology by Zhang (2003). Transaction costs for S\&P 500 are set equal to 0.005\%. (Michańków et al., 2022)}
\end{minipage}}
\label{fig:equity_1}
\end{figure}

The performance, both in terms of error metrics and trading indicators, of econometric techniques, machine learning models, and hybrid approaches for the S\&P 500 index and \textit{Long-Short} trading strategy is presented in Table \ref{tab:equity_1}. Analogically, Figure \ref{fig:equity_1} illustrates the equity lines for the aforementioned models. The out-of-sample part of the data for S\&P 500 spanned the period from 1 January 2007 to 31 December 2023. 

The most accurate predictions, based on the RMSE and MAE metrics, were generated by the SVM-ARFIMA (1) model. In addition, SVM-ARIMA (1), LSTM-ARIMA (1), and LSTM-ARFIMA (1) achieved superior results compared to their individual components. Two conclusions can be drawn. First, XGBoost and the XGBoost-based hybrid models failed to deliver high-quality forecasts for the S\&P 500 index. Secondly, the hybridization methodology by Zhang (2003) provided inferior predictions in comparison to the hybrid method of inputting a statistical model forecast as an additional explanatory feature to the machine learning model.

Evaluating the models' trading performance based on the three risk-adjusted return metrics, the individual SVM model served as the best-performing investment strategy. With the IR metric at the level of 0.68, SVM significantly outperformed the benchmark Buy\&Hold. The following techniques also beat the market: ARIMA, ARFIMA, SVM-ARIMA (1), SVM-ARFIMA(1), SVM-ARIMA(2), LSTM-ARIMA(1), and LSTM-ARFIMA (2). Once again, none of the XGBoost-based models delivered satisfactory performance. This observation may support the findings by Lv et al. (2019), who demonstrated the trading performance degradation for the XGBoost model after the incorporation of transaction costs.

\begin{figure}[H]
\centering
\caption{Equity lines for S\&P 500 and \textit{Long Only} trading strategy}
\includegraphics[width=1\textwidth]{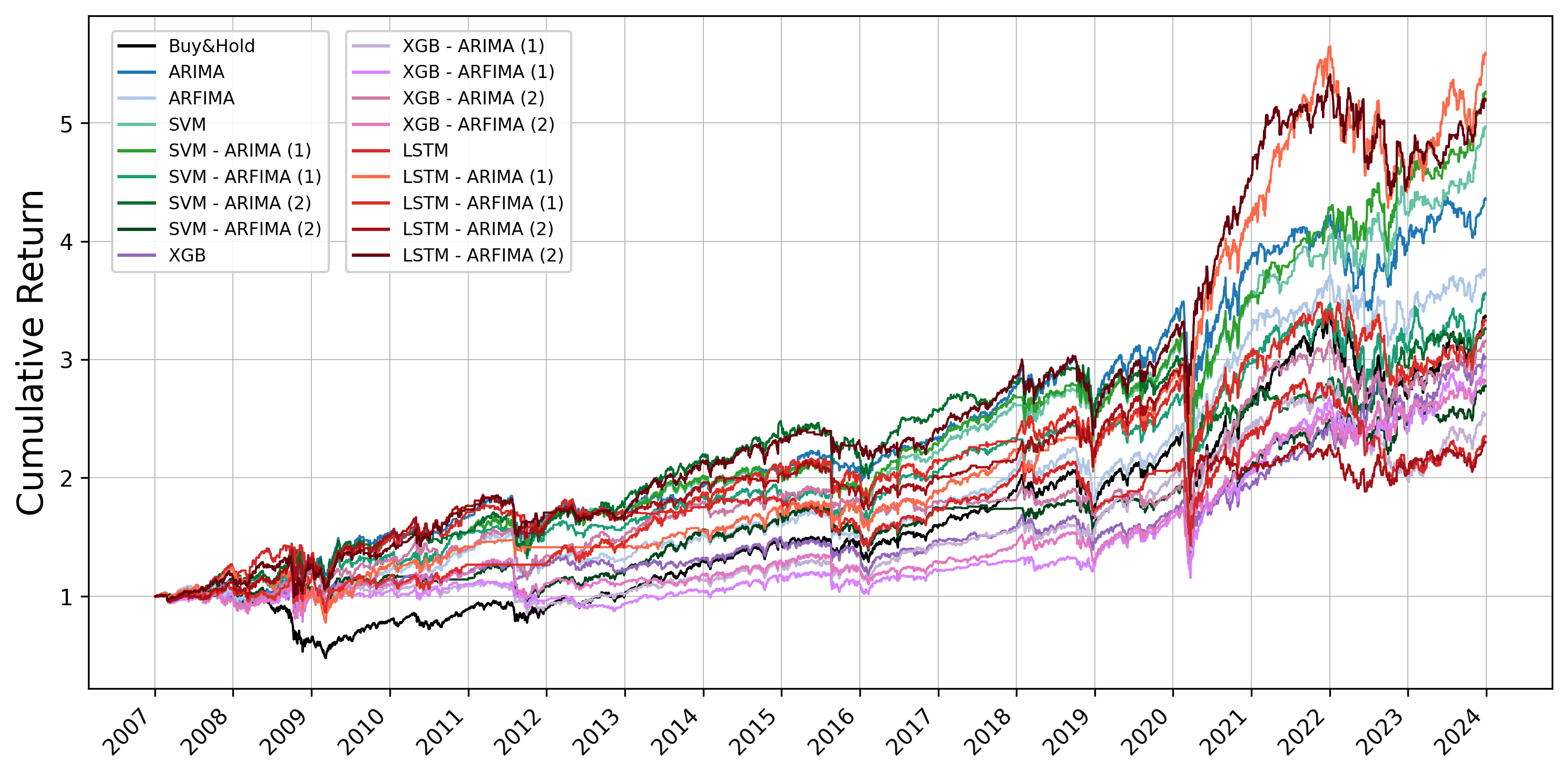}
\caption*{\begin{minipage}[t]{\textwidth}
\raggedright \scriptsize\textit{Note: All equity lines refer to predictive models out-of-sample forecasts. The first line represents the benchmark Buy\&Hold strategy. Annotation (1) denotes the hybridization technique of inputting the prediction of the econometric model to the machine learning model, while (2) indicates the hybrid methodology by Zhang (2003). Transaction costs for S\&P 500 are set equal to 0.005\%. (Michańków et al., 2022)}
\end{minipage}}
\label{fig:equity_2}
\end{figure}

Table \ref{tab:equity_2} and Figure \ref{fig:equity_2} contain the performance of the compared models for S\&P 500 and \textit{Long Only} signals. Based on the IR, IR*, and SR metrics, the best results were achieved by the LSTM-ARFIMA (2), LSTM-ARIMA (1), and SVM-ARIMA (1) models. Moreover, the annual rate of return for these techniques exceeded 10\% significantly outperforming the Buy\&Hold scenario. These hybrid models also delivered superior performance compared to their individual components. It should be noted that all predictive model-based strategies are characterized by notably lower levels of annualized standard deviation and maximum drawdown relative to the market.

\begin{table}[htbp]
\begin{center}
\caption{Forecasting performance of multiple models for S\&P 500 and \textit{Long-Short} trading strategy}
\label{tab:equity_1}
{\scriptsize
\begin{tabular}{p{2.8cm} p{0.9cm} p{0.9cm} p{0.7cm} p{0.7cm} p{0.7cm} p{0.7cm} p{0.7cm} p{0.7cm}}
\toprule
\textbf{Method} & \multicolumn{2}{c}{\textbf{Error metrics}} & \multicolumn{6}{c}{\textbf{Performance indicators}} \\
\cmidrule(lr){2-3} \cmidrule(lr){4-9}
 & RMSE & MAE & ARC & ASD  & MD & IR & IR* & SR \\
\midrule
Buy\&Hold S\&P & - & - & 7.39\% & 20.27\% & 56.78\% & 0.36 & 0.05 & 0.56  \\
ARIMA & 1.2890\% & 0.8322\% & 6.89\% & \textbf{14.24\%} & 30.21\% & 0.48 & 0.11 & 0.81 \\
ARFIMA & 1.2858\% & 0.8284\% & 7.39\% & 16.63\% & \textbf{27.18\%} & 0.44 & 0.12 & 0.70 \\
SVM & 1.2812\% & 0.8253\% & \textbf{12.20\%} & 18.00\% & 33.92\% & \textbf{0.68} & \textbf{0.24} & \textbf{1.09} \\
SVM-ARIMA (1) & 1.2812\% & 0.8249\% & 11.96\% & 18.08\% & 33.92\% & 0.66 & 0.23 & 1.06 \\
SVM-ARFIMA (1) & \textbf{1.2811\%} & \textbf{0.8229\%} & 8.11\% & 17.86\% & 31.02\% & 0.45 & 0.12 & 0.73 \\
SVM-ARIMA (2) & 1.3065\% & 0.8489\% & 6.87\% & 17.29\% & 35.43\% & 0.40 & 0.08 & 0.65 \\
SVM-ARFIMA (2) & 1.3007\% & 0.8485\% & 6.01\% & 17.46\% & 37.88\% & 0.34 & 0.05 & 0.55 \\
XGBoost & 1.2913\% & 0.8292\% & 3.37\% & 18.00\% & 47.84\% & 0.19 & 0.01 & 0.29 \\
XGBoost-ARIMA (1) & 1.2983\% & 0.8342\% & 4.65\% & 17.95\% & 42.24\% & 0.26 & 0.03 & 0.40 \\
XGBoost-ARFIMA (1) & 1.2863\% & 0.8275\% & 3.83\% & 18.18\% & 56.47\% & 0.21 & 0.01 & 0.33 \\
XGBoost-ARIMA (2) & 1.3075\% & 0.8492\% & 2.71\% & 17.26\% & 35.80\% & 0.16 & 0.01 & 0.25 \\
XGBoost-ARFIMA (2) & 1.2973\% & 0.8378\% & 4.05\% & 16.59\% & 31.25\% & 0.24 & 0.03 & 0.38 \\
LSTM & 1.2881\% & 0.8259\% & 0.50\% & 18.42\% & 65.84\% & 0.03 & 0.00 & 0.04 \\
LSTM-ARIMA (1) & 1.2825\% & 0.8226\% & 10.46\% & 18.84\% & 36.58\% & 0.56 & 0.16 & 0.89 \\
LSTM-ARFIMA (1) & 1.2832\% & 0.8232\% & 6.31\% & 18.60\% & 33.92\% & 0.34 & 0.06 & 0.54 \\
LSTM-ARIMA (2) & 1.3114\% & 0.8469\% & 5.01\% & 17.40\% & 43.67\% & 0.29 & 0.03 & 0.45 \\
LSTM-ARFIMA (2) & 1.2934\% & 0.8392\% & 8.87\% & 17.17\% & 33.25\% & 0.52 & 0.14 & 0.82 \\
\bottomrule
\multicolumn{9}{p{12cm}}{\raggedright\scriptsize\textit{Note: All values refer to performance indicators derived from predictive models out-of-sample forecasts. The first column represents the benchmark Buy\&Hold strategy. Annotation (1) denotes the hybridization technique of inputting the prediction of the econometric model to the machine learning model, while (2) indicates the hybrid methodology by Zhang (2003).}}
\end{tabular}
}
\end{center}
\end{table}

\begin{table}[htbp]
\begin{center}
\scriptsize
\caption{Forecasting performance of multiple models for S\&P 500 and \textit{Long Only} trading strategy}
\label{tab:equity_2}
\begin{tabular}{p{3.2cm} p{1.1cm} p{1.1cm} p{1.1cm} p{1.1cm} p{1.1cm} p{1.1cm}}
\toprule
\textbf{Method} & \multicolumn{6}{c}{\textbf{Performance indicators}} \\
\cmidrule(lr){2-7}
 & ARC & ASD & MD & IR & IR* & SR \\
\midrule
Buy\&Hold S\&P 500 & 7.40\% & 20.27\% & 56.78\% & 0.36 & 0.05 & 0.56 \\
ARIMA & 9.04\% & 16.06\% & 33.92\% & 0.56 & 0.15 & 0.86 \\
ARFIMA & 8.10\% & 16.06\% & 29.56\% & 0.50 & 0.14 & 0.76 \\
SVM & 9.87\% & 16.63\% & 33.92\% & 0.59 & 0.17 & 0.91 \\
SVM-ARIMA (1) & 10.25\% & 16.67\% & 33.92\% & 0.61 & 0.19 & 0.94 \\
SVM-ARFIMA (1) & 7.75\% & 16.66\% & 30.67\% & 0.47 & 0.12 & 0.71 \\
SVM-ARIMA (2) & 7.19\% & \textbf{15.88\%} & 33.56\% & 0.45 & 0.10 & 0.69 \\
SVM-ARFIMA (2) & 6.18\% & 16.24\% & 32.00\% & 0.38 & 0.07 & 0.57 \\
XGBoost & 6.71\% & 17.27\% & 31.06\% & 0.39 & 0.08 & 0.59 \\
XGBoost-ARIMA (1) & 5.62\% & 17.05\% & 33.92\% & 0.33 & 0.05 & 0.50 \\
XGBoost-ARFIMA (1) & 6.56\% & 16.88\% & 34.12\% & 0.39 & 0.07 & 0.59 \\
XGBoost-ARIMA (2) & 6.99\% & 16.17\% & 27.57\% & 0.43 & 0.11 & 0.67 \\
XGBoost-ARFIMA (2) & 6.29\% & 15.69\% & 29.56\% & 0.40 & 0.09 & 0.60 \\
LSTM & 5.15\% & 16.03\% & 33.92\% & 0.32 & 0.05 & 0.48 \\
LSTM-ARIMA (1) & \textbf{10.64\%} & 17.48\% & 37.47\% & 0.61 & 0.17 & 0.94 \\
LSTM-ARFIMA (1) & 7.33\% & 16.92\% & 33.92\% & 0.43 & 0.09 & 0.66 \\
LSTM-ARIMA (2) & 5.02\% & 15.85\% & 38.50\% & 0.32 & 0.04 & 0.47 \\
LSTM-ARFIMA (2) & 10.17\% & 16.09\% & \textbf{25.56\%} & \textbf{0.63} & \textbf{0.25} & \textbf{0.97} \\
\bottomrule
\multicolumn{7}{p{12cm}}{\raggedright\scriptsize\textit{Note: All values refer to performance indicators derived from predictive models out-of-sample forecasts. The first row represents the benchmark Buy\&Hold strategy. Annotation (1) denotes the hybridization technique of inputting the prediction of the econometric model to the machine learning model, while (2) indicates the hybrid methodology by Zhang (2003).}}
\end{tabular}
\end{center}
\end{table}

\subsection{Investment strategies for Bitcoin}

\begin{figure}[H]
\centering
\caption{Equity lines for Bitcoin and \textit{Long-Short} trading strategy}
\includegraphics[width=1\textwidth]{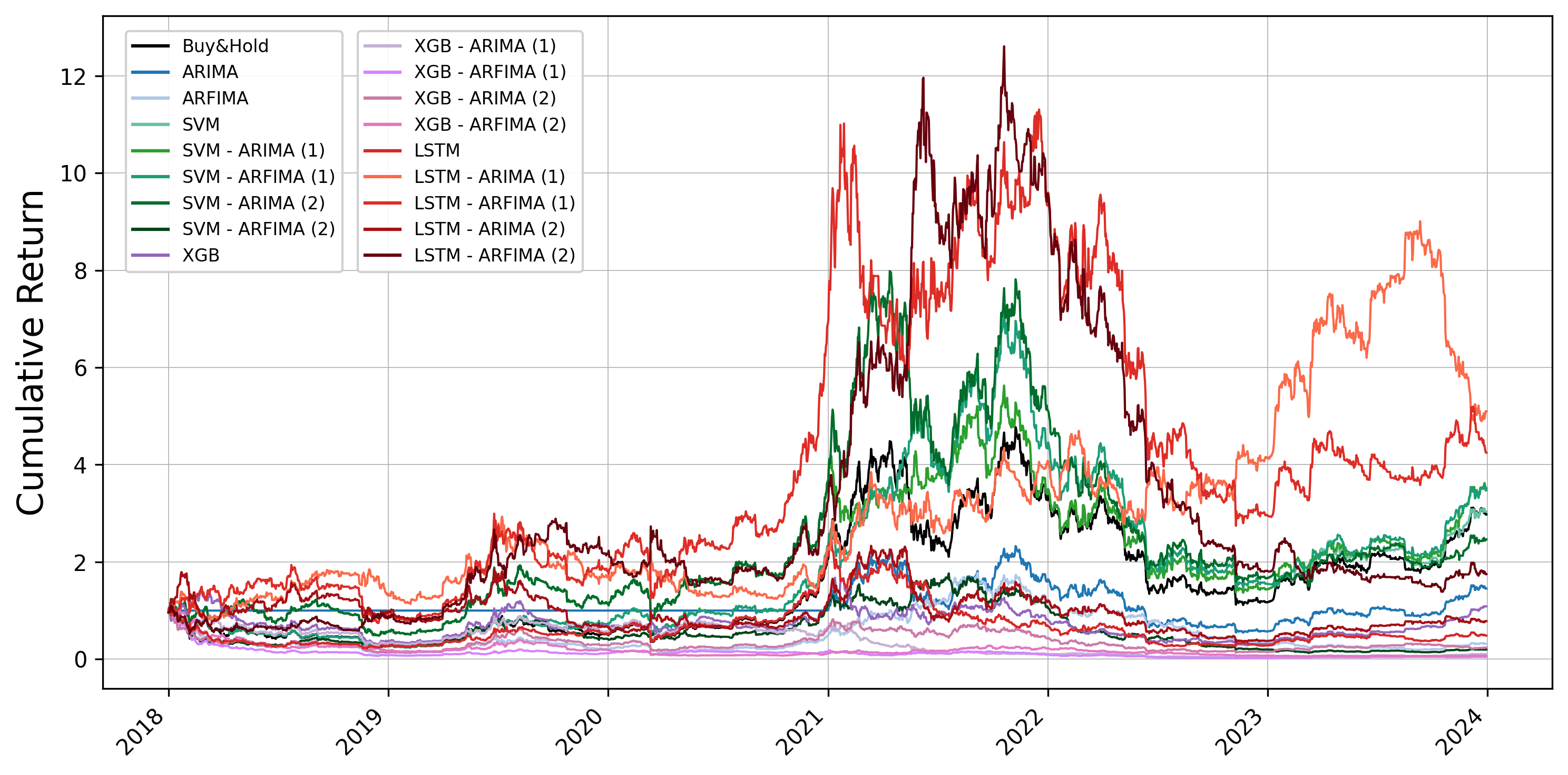}
\caption*{\begin{minipage}[t]{\textwidth}
\raggedright \scriptsize\textit{Note: All equity lines refer to predictive models out-of-sample forecasts. The first line represents the benchmark Buy\&Hold strategy. Annotation (1) denotes the hybridization technique of inputting the prediction of the econometric model to the machine learning model, while (2) indicates the hybrid methodology by Zhang (2003). Transaction costs for Bitcoin are set equal to 0.01\%. (Michańków et al., 2022)}
\end{minipage}}
\label{fig:equity_3}
\end{figure}

The out-of-sample period for Bitcoin consists of daily observations between 1 January 2018 and 31 December 2023. Table \ref{tab:equity_3} reports the computed values of forecasting error metrics and performance indicators for the \textit{Long-Short} trading strategy. Correspondingly, Figure \ref{fig:equity_3} illustrates the investment levels and dynamics of all analyzed models.

The most accurate predictions, according to the RMSE and MAE metrics, were generated by the ARIMA model. Relatively good prediction accuracy was also achieved by the ARFIMA and SVM-based models. However, in the case of Bitcoin data, combining the models did not lead to improved predictive power. These findings support the results presented by Dudek et al. (2024). Additionally, the econometric models outperformed the machine learning techniques in terms of forecasting accuracy. 

More promising results in favor of hybridization were achieved during the backtesting procedure. \textit{Long-Short} trading signals applied to the predictions of LSTM-ARIMA (1) resulted in IR and SR metrics at the levels of 0.50 and 1.06, respectively. The SVM-ARIMA (1), SVM-ARIMA (2), and LSTM-ARFIMA (1) also managed to outperform both the market and their individual underlying models. These observations are consistent with the findings for the S\&P 500 index: the unsuitability of XGBoost-based approaches and the superior performance of hybrid methodology not assuming an additive relationship between linear and nonlinear components. It is noteworthy that the ASD and MD values for Bitcoin are significantly higher compared to those for the S\&P 500 index. It supports the view of cryptocurrency markets as more chaotic and harder to predict.

\begin{figure}[H]
\centering
\caption{Equity lines for Bitcoin and \textit{Long Only} trading strategy}
\includegraphics[width=1\textwidth]{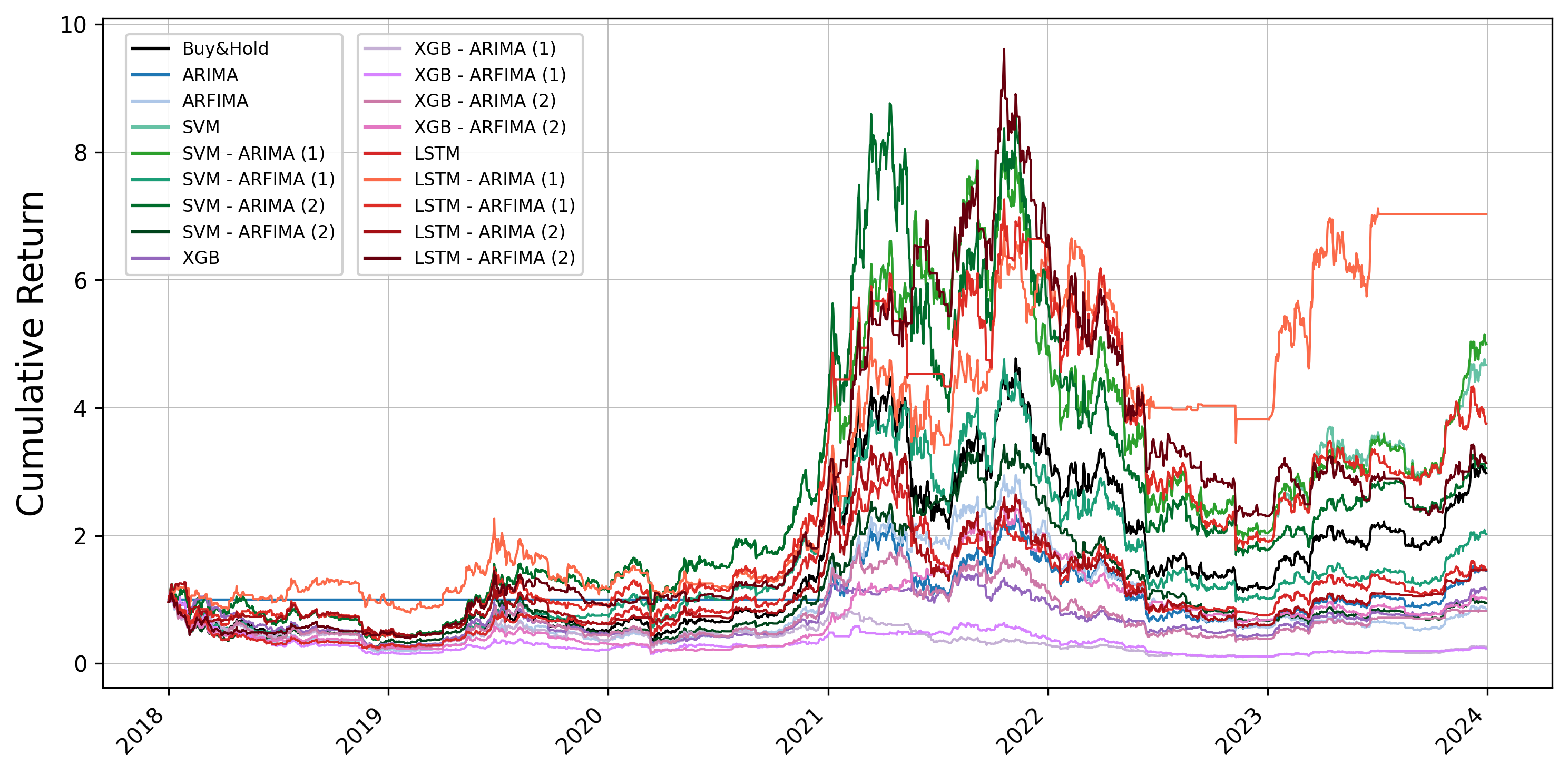}
\caption*{\begin{minipage}[t]{\textwidth}
\raggedright \scriptsize\textit{Note: All equity lines refer to predictive models out-of-sample forecasts. The first line represents the benchmark Buy\&Hold strategy. Annotation (1) denotes the hybridization technique of inputting the prediction of the econometric model to the machine learning model, while (2) indicates the hybrid methodology by Zhang (2003). Transaction costs for Bitcoin are set equal to 0.01\%. (Michańków et al., 2022)}
\end{minipage}}
\label{fig:equity_4}
\end{figure}

Table \ref{tab:equity_4} and Figure \ref{fig:equity_4} summarize the performance of various applied approaches for \textit{Long Only} signals. With regard to econometric models, ARIMA outperformed the ARFIMA alternative. However, neither was able to beat the market. Among machine learning techniques, only the SVM model achieved superior return, risk, and risk-adjusted return metrics compared to the Buy\&Hold benchmark. From the full set of evaluated models, the most suitable investment strategy was constructed using the hybrid LSTM-ARIMA (1) model. Hybridization outperformed its individual components also in the cases of SVM-ARIMA (1), LSTM-ARFIMA (1), and LSTM-ARFIMA (2).

\begin{table}[htbp]
\begin{center}
\caption{Forecasting performance of multiple models for Bitcoin and \textit{Long-Short} trading strategy}
\label{tab:equity_3}
{\scriptsize
\begin{tabular}{p{2.8cm} p{0.9cm} p{0.9cm} p{1cm} p{0.7cm} p{0.7cm} p{0.7cm} p{0.7cm} p{0.7cm}}
\toprule
\textbf{Method} & \multicolumn{2}{c}{\textbf{Error metrics}} & \multicolumn{6}{c}{\textbf{Performance indicators}} \\
\cmidrule(lr){2-3} \cmidrule(lr){4-9}
 & RMSE & MAE & ARC & ASD  & MD & IR & IR* & SR \\
\midrule
Buy\&Hold Bitcoin & - & - & 19.95\% & 69.43\% & 81.53\% & 0.29 & 0.07 & 0.56  \\
ARIMA & \textbf{3.6858\%} & \textbf{2.4229\%} & 6.41\% & \textbf{45.55\%} & 76.63\% & 0.14 & 0.01 & 0.26 \\
ARFIMA & 3.7089\% & 2.4383\% & -17.48\% & 63.04\% & 89.39\% & -0.28 & -0.05 & -0.51 \\
SVM & 3.6940\% & 2.4418\% & 20.32\% & 63.04\% & 75.72\% & 0.32 & 0.09 & 0.61 \\
SVM-ARIMA (1) & 3.6928\% & 2.4448\% & 23.32\% & 62.99\% & 75.72\% & 0.37 & 0.11 & 0.71 \\
SVM-ARFIMA (1) & 3.7093\% & 2.4589\% & 23.07\% & 64.74\% & 81.89\% & 0.36 & 0.10 & 0.67 \\
SVM-ARIMA (2) & 3.7075\% & 2.4550\% & 16.22\% & 64.47\% & 81.27\% & 0.25 & 0.05 & 0.48 \\
SVM-ARFIMA (2) & 3.7278\% & 2.4697\% & -23.71\% & 58.59\% & 92.67\% & -0.40 & -0.10 & -0.76 \\
XGBoost & 3.7095\% & 2.4600\% & 1.38\% & 57.48\% & 84.07\% & 0.02 & 0.00 & 0.05 \\
XGBoost-ARIMA (1) & 3.7354\% & 2.4736\% & -31.62\% & 63.93\% & 96.59\% & -0.49 & -0.16 & -0.89 \\
XGBoost-ARFIMA (1) & 3.7819\% & 2.5025\% & -40.58\% & 61.27\% & 97.61\% & -0.66 & -0.28 & -1.20 \\
XGBoost-ARIMA (2) & 3.7144\% & 2.4555\% & -21.22\% & 61.82\% & 86.46\% & -0.34 & -0.08 & -0.64 \\
XGBoost-ARFIMA (2) & 3.7444\% & 2.4817\% & -35.93\% & 58.37\% & 94.40\% & -0.62 & -0.23 & -1.09 \\
LSTM & 3.7570\% & 2.5007\% & -11.36\% & 61.95\% & 89.17\% & -0.18 & -0.02 & -0.34 \\
LSTM-ARIMA (1) & 3.7249\% & 2.4786\% & \textbf{31.19\%} & 61.83\% & \textbf{57.71\%} & \textbf{0.50} & \textbf{0.27} & \textbf{1.06} \\
LSTM-ARFIMA (1) & 3.7318\% & 2.4831\% & 27.26\% & 64.71\% & 75.74\% & 0.42 & 0.15 & 0.81 \\
LSTM-ARIMA (2) & 3.7362\% & 2.4909\% & -3.98\% & 65.21\% & 84.72\% & -0.06 & 0.00 & -0.12 \\
LSTM-ARFIMA (2) & 3.7362\% & 2.4799\% & 9.76\% & 62.08\% & 88.99\% & 0.16 & 0.02 & 0.31 \\
\bottomrule
\multicolumn{9}{p{12cm}}{\raggedright\scriptsize\textit{Note: All values refer to performance indicators derived from predictive models out-of-sample forecasts. The first column represents the benchmark Buy\&Hold strategy. Annotation (1) denotes the hybridization technique of inputting the prediction of the econometric model to the machine learning model, while (2) indicates the hybrid methodology by Zhang (2003).}}
\end{tabular}
}
\end{center}
\end{table}

\begin{table}[htbp]
\begin{center}
\scriptsize
\caption{Forecasting performance of multiple models for Bitcoin and \textit{Long Only} trading strategy}
\label{tab:equity_4}
\begin{tabular}{p{3.2cm} p{1.1cm} p{1.1cm} p{1.1cm} p{1.1cm} p{1.1cm} p{1.1cm}}
\toprule
\textbf{Method} & \multicolumn{6}{c}{\textbf{Performance indicators}} \\
\cmidrule(lr){2-7}
 & ARC & ASD & MD & IR & IR* & SR \\
\midrule
Buy\&Hold Bitcoin & 19.95\% & 69.43\% & 81.53\% & 0.29 & 0.07 & 0.56 \\
ARIMA & 6.41\% & \textbf{45.55\%} & 76.63\% & 0.14 & 0.01 & 0.26 \\
ARFIMA & -2.52\% & 61.96\% & 83.64\% & -0.04 & 0.00 & -0.08 \\
SVM & 29.28\% & 63.39\% & 76.45\% & 0.46 & 0.18 & 0.89 \\
SVM-ARIMA (1) & 30.76\% & 63.44\% & 76.45\% & 0.48 & 0.20 & 0.94 \\
SVM-ARFIMA (1) & 12.50\% & 63.74\% & 81.14\% & 0.20 & 0.03 & 0.37 \\
SVM-ARIMA (2) & 20.48\% & 63.87\% & 80.74\% & 0.32 & 0.08 & 0.61 \\
SVM-ARFIMA (2) & -0.96\% & 58.70\% & 82.03\% & -0.02 & 0.00 & -0.03 \\
XGBoost & 2.51\% & 54.06\% & 77.02\% & 0.05 & 0.00 & 0.08 \\
XGBoost-ARIMA (1) & -20.12\% & 60.14\% & 91.34\% & -0.33 & -0.07 & -0.61 \\
XGBoost-ARFIMA (1) & -21.45\% & 57.64\% & 90.79\% & -0.37 & -0.09 & -0.68 \\
XGBoost-ARIMA (2) & -3.18\% & 61.96\% & 81.57\% & -0.05 & 0.00 & -0.10 \\
XGBoost-ARFIMA (2) & 0.20\% & 54.14\% & 84.96\% & 0.00 & 0.00 & 0.01 \\
LSTM & 6.64\% & 59.11\% & 81.03\% & 0.11 & 0.01 & 0.21 \\
LSTM-ARIMA (1) & \textbf{38.40\%} & 53.19\% & \textbf{52.80\%} & \textbf{0.72} & \textbf{0.53} & \textbf{1.43} \\
LSTM-ARFIMA (1) & 24.64\% & 60.16\% & 75.55\% & 0.41 & 0.13 & 0.77 \\
LSTM-ARIMA (2) & 6.49\% & 58.61\% & 83.39\% & 0.11 & 0.01 & 0.21 \\
LSTM-ARFIMA (2) & 21.01\% & 54.05\% & 77.55\% & 0.39 & 0.11 & 0.76 \\
\bottomrule
\multicolumn{7}{p{12cm}}{\raggedright\scriptsize\textit{Note: All values refer to performance indicators derived from predictive models out-of-sample forecasts. The first row represents the benchmark Buy\&Hold strategy. Annotation (1) denotes the hybridization technique of inputting the prediction of the econometric model to the machine learning model, while (2) indicates the hybrid methodology by Zhang (2003).}}
\end{tabular}
\end{center}
\end{table}

\subsection{Investment strategies for the portfolio of assets}

Asset allocation and portfolio optimization represent another broad area of research in the field of financial forecasting (Malandri et al., 2018; Ma et al., 2020; Ma et al., 2021). Due to its relative simplicity, the equally weighted method of portfolio construction for the S\&P 500 index and Bitcoin is applied in this study. The construction of the portfolio aims to minimize asset-specific fluctuations and provide a more holistic view on each model's performance evaluation. The backtesting period spanned from 1 January 2018 to 31 December 2023.

\begin{figure}[H]
\centering
\caption{Equity lines for the portfolio of assets and \textit{Long-Short} trading strategy}
\includegraphics[width=1\textwidth]{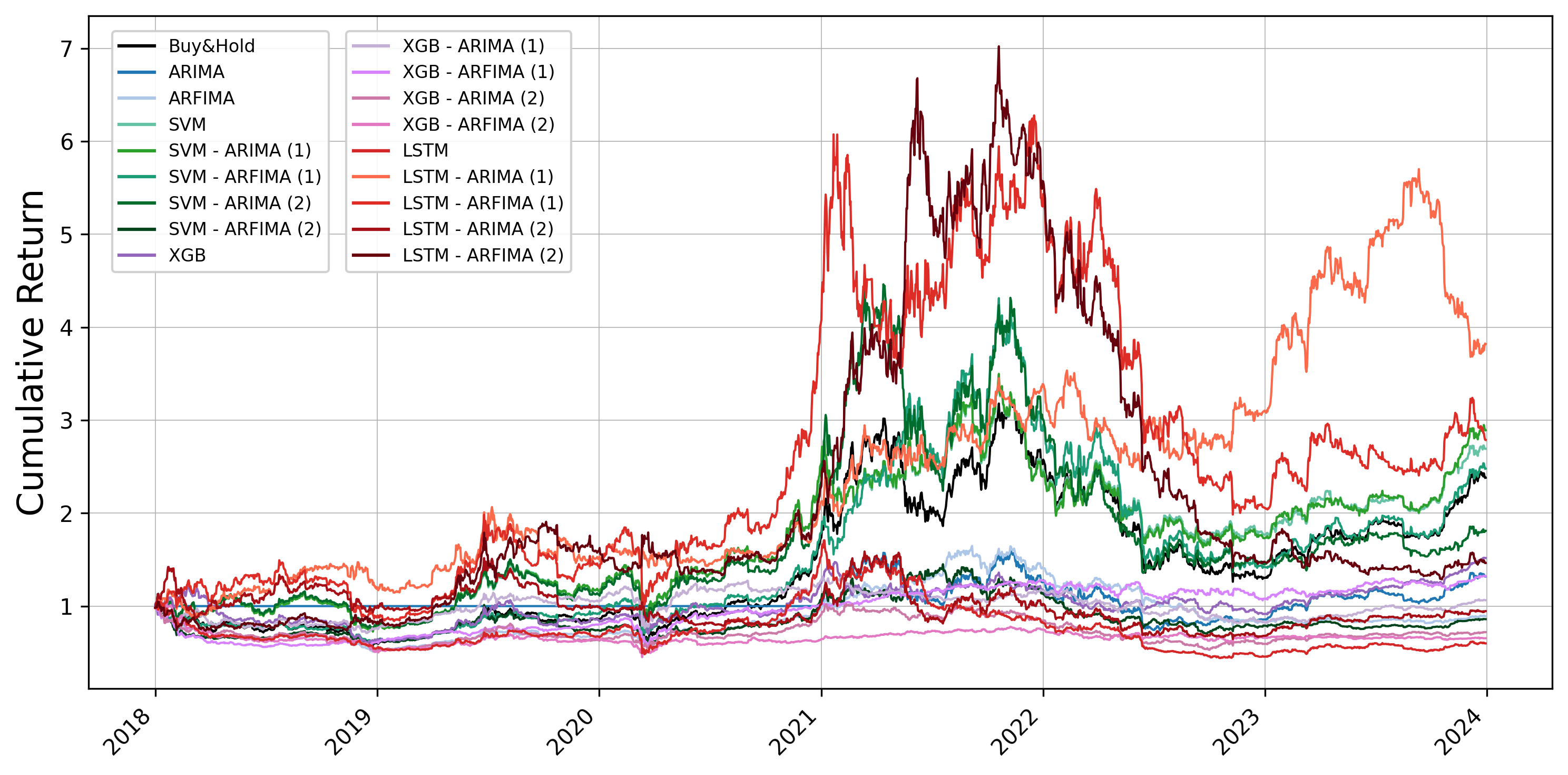}
\caption*{\begin{minipage}[t]{\textwidth}
\raggedright \scriptsize\textit{Note: All equity lines refer to predictive models out-of-sample forecasts. The first line represents the benchmark Buy\&Hold strategy. Annotation (1) denotes the hybridization technique of inputting the prediction of the econometric model to the machine learning model, while (2) indicates the hybrid methodology by Zhang (2003). Transaction costs for S\&P 500 are set equal to 0.0005\%, and for Bitcoin to 0.01\%. (Michańków et al., 2022)}
\end{minipage}}
\label{fig:equity_5}
\end{figure}

For the \textit{Long-Short} strategy, Table \ref{tab:equity_5} and Figure \ref{fig:equity_5} illustrate the trading outcomes and corresponding investment indicators for all evaluated models. Significantly superior effectiveness was demonstrated by the hybrid LSTM-ARIMA (1) model, achieving an annual rate of return exceeding 25\%. The risk-adjusted return metrics further support this finding. The outperformance of the market was also observed for the SVM, SVM-ARIMA (1), and SVM-ARFIMA (1) models. However, only the LSTM-ARIMA (1) and SVM-ARIMA(1) techniques managed to yielded improvements over their individual components. This observation supports previous findings regarding the suitability of ARIMA, SVM, and LSTM models, particularly when combined through a non-additive hybridization approach, for constructing profitable investment strategies.

\begin{figure}[H]
\centering
\caption{Equity lines for the portfolio of assets and \textit{Long Only} trading strategy}
\includegraphics[width=1\textwidth]{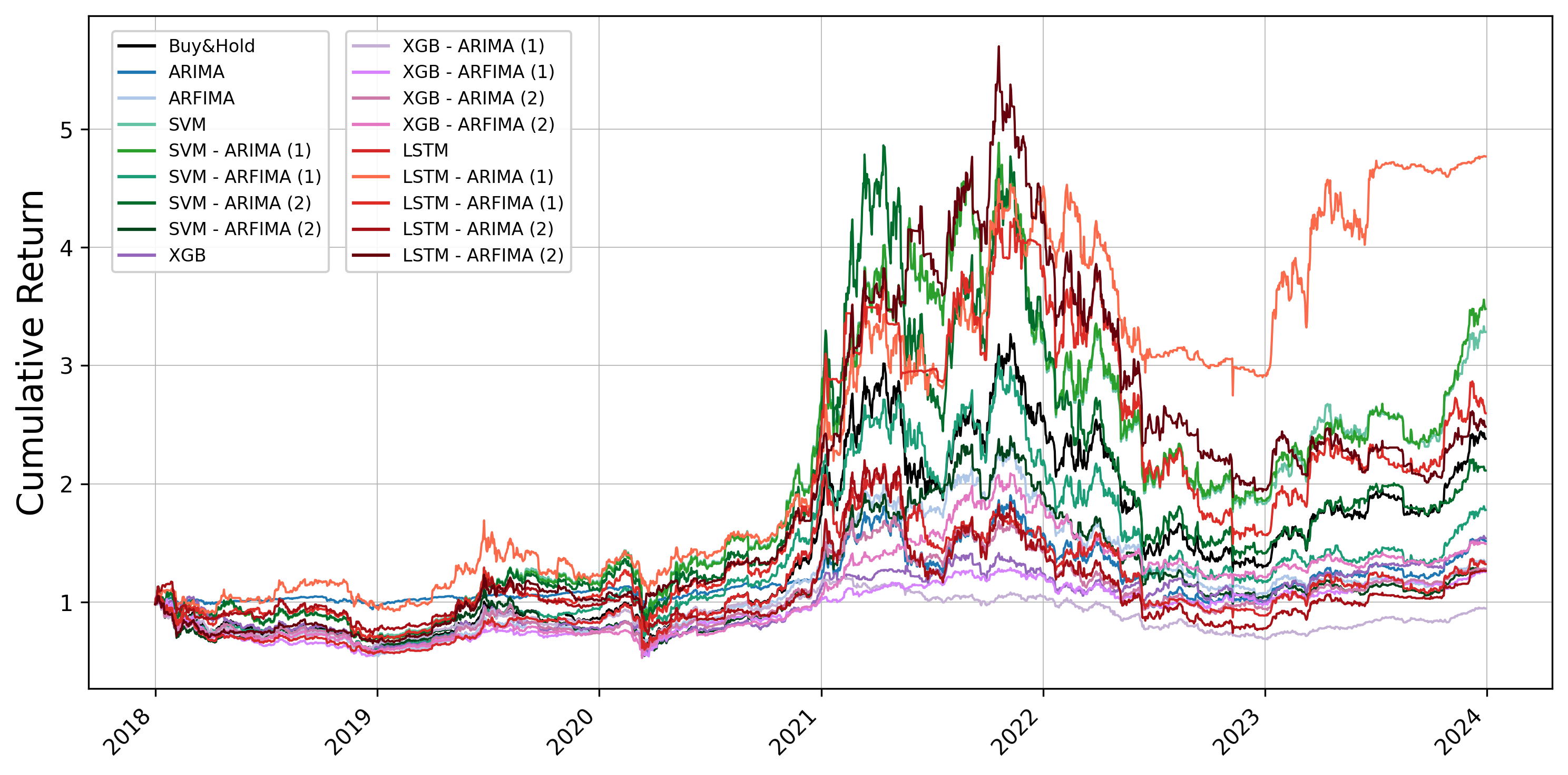}
\caption*{\begin{minipage}[t]{\textwidth}
\raggedright \scriptsize\textit{Note: All equity lines refer to predictive models out-of-sample forecasts. The first line represents the benchmark Buy\&Hold strategy. Annotation (1) denotes the hybridization technique of inputting the prediction of the econometric model to the machine learning model, while (2) indicates the hybrid methodology by Zhang (2003). Transaction costs for S\&P 500 are set equal to 0.0005\%, and for Bitcoin to 0.01\%. (Michańków et al., 2022)}
\end{minipage}}
\label{fig:equity_6}
\end{figure}

Analogically, Table \ref{tab:equity_6} and Figure \ref{fig:equity_6} contain the models' trading performance based on the \textit{Long Only} signals. Once again, the hybrid LSTM-ARIMA (1) model was characterized by the most effective trading performance, even outperforming the previous strategy based on the \textit{Long-Short} signals. The annualized rate of return exceeded the level of 30\%, and the IR metric reached 0.91. Notably, the maximum drawdown was significantly lower than that of other approaches, indicating a higher degree of resilience to volatile market conditions. Among the individual models, only the SVM managed to outperform the market. In the case of hybrid models, positive results were also reported for SVM-ARIMA (1), LSTM-ARFIMA (1), and LSTM-ARFIMA(2). These findings are consistent with previous conclusions.

\begin{table}[htbp]
\begin{center}
\scriptsize
\caption{Forecasting performance of multiple models for the portfolio of assets and \textit{Long-Short} trading strategy}
\label{tab:equity_5}
\begin{tabular}{p{3.2cm} p{1.1cm} p{1.1cm} p{1.1cm} p{1.1cm} p{1.1cm} p{1.1cm}}
\toprule
\textbf{Method} & \multicolumn{6}{c}{\textbf{Performance indicators}} \\
\cmidrule(lr){2-7}
 & ARC & ASD & MD & IR & IR* & SR \\
\midrule
Buy\&Hold portfolio & 15.90\% & 38.45\% & 61.36\% & 0.41 & 0.11 & 0.78 \\
ARIMA & 4.71\% & 25.95\% & 54.20\% & 0.18 & 0.02 & 0.33 \\
ARFIMA & -1.61\% & 23.86\% & 52.24\% & -0.07 & 0.00 & -0.12 \\
SVM & 18.31\% & 37.81\% & 52.25\% & 0.48 & 0.17 & 0.91 \\
SVM-ARIMA (1) & 19.72\% & 37.88\% & 53.31\% & 0.52 & 0.19 & 0.98 \\
SVM-ARFIMA (1) & 16.69\% & 37.97\% & 68.31\% & 0.44 & 0.11 & 0.82 \\
SVM-ARIMA (2) & 10.73\% & 43.06\% & 69.57\% & 0.25 & 0.04 & 0.47 \\
SVM-ARFIMA (2) & -2.19\% & 25.79\% & 51.42\% & -0.09 & 0.00 & -0.16 \\
XGBoost & 6.89\% & 26.00\% & 50.75\% & 0.27 & 0.04 & 0.50 \\
XGBoost-ARIMA (1) & 1.02\% & 23.77\% & 38.16\% & 0.04 & 0.00 & 0.07 \\
XGBoost-ARFIMA (1) & 5.08\% & 19.85\% & 42.67\% & 0.26 & 0.03 & 0.45 \\
XGBoost-ARIMA (2) & -5.04\% & 21.79\% & 48.06\% & -0.23 & -0.02 & -0.42 \\
XGBoost-ARFIMA (2) & -6.54\% & \textbf{18.39\%} & 56.43\% & -0.36 & -0.04 & -0.61 \\
LSTM & -7.91\% & 30.82\% & 74.16\% & -0.26 & -0.03 & -0.46 \\
LSTM-ARIMA (1) & \textbf{25.41\%} & 36.94\% & \textbf{35.43\%} & \textbf{0.69} & \textbf{0.49} & \textbf{1.42} \\
LSTM-ARFIMA (1) & 18.99\% & 48.38\% & 68.38\% & 0.39 & 0.11 & 0.75 \\
LSTM-ARIMA (2) & -0.62\% & 34.48\% & 59.33\% & -0.02 & 0.00 & -0.03 \\
LSTM-ARFIMA (2) & 6.88\% & 42.20\% & 81.66\% & 0.16 & 0.01 & 0.32 \\
\bottomrule
\multicolumn{7}{p{12cm}}{\raggedright\scriptsize\textit{Note: All values refer to performance indicators derived from predictive models out-of-sample forecasts. The first row represents the benchmark Buy\&Hold strategy. Annotation (1) denotes the hybridization technique of inputting the prediction of the econometric model to the machine learning model, while (2) indicates the hybrid methodology by Zhang (2003).}}
\end{tabular}
\end{center}
\end{table}

\begin{table}[htbp]
\begin{center}
\scriptsize
\caption{Forecasting performance of multiple models for the portfolio of assets and \textit{Long Only} trading strategy}
\label{tab:equity_6}
\begin{tabular}{p{3.2cm} p{1.1cm} p{1.1cm} p{1.1cm} p{1.1cm} p{1.1cm} p{1.1cm}}
\toprule
\textbf{Method} & \multicolumn{6}{c}{\textbf{Performance indicators}} \\
\cmidrule(lr){2-7}
 & ARC & ASD & MD & IR & IR* & SR \\
\midrule
Buy\&Hold portfolio & 15.90\% & 38.45\% & 61.36\% & 0.41 & 0.11 & 0.78 \\
ARIMA & 7.17\% & 24.27\% & 49.87\% & 0.30 & 0.04 & 0.55 \\
ARFIMA & 5.33\% & 27.80\% & 54.81\% & 0.19 & 0.02 & 0.35 \\
SVM & 22.28\% & 40.71\% & 63.80\% & 0.55 & 0.19 & 1.04 \\
SVM-ARIMA (1) & 23.47\% & 40.77\% & 63.41\% & 0.58 & 0.21 & 1.10 \\
SVM-ARFIMA (1) & 10.40\% & 36.38\% & 63.30\% & 0.29 & 0.05 & 0.53 \\
SVM-ARIMA (2) & 13.61\% & 43.39\% & 72.24\% & 0.31 & 0.06 & 0.59 \\
SVM-ARFIMA (2) & 4.34\% & 29.66\% & 58.93\% & 0.15 & 0.01 & 0.26 \\
XGBoost & 7.45\% & 24.73\% & 45.85\% & 0.30 & 0.05 & 0.53 \\
XGBoost-ARIMA (1) & -0.93\% & 23.33\% & 43.90\% & -0.04 & 0.00 & -0.07 \\
XGBoost-ARFIMA (1) & 4.17\% & \textbf{21.61\%} & 48.13\% & 0.19 & 0.02 & 0.34 \\
XGBoost-ARIMA (2) & 4.54\% & 26.43\% & 47.88\% & 0.17 & 0.02 & 0.31 \\
XGBoost-ARFIMA (2) & 7.21\% & 22.96\% & 49.97\% & 0.31 & 0.05 & 0.56 \\
LSTM & 5.06\% & 32.07\% & 59.81\% & 0.16 & 0.01 & 0.29 \\
LSTM-ARIMA (1) & \textbf{30.13\%} & 33.26\% & \textbf{39.99\%} & \textbf{0.91} & \textbf{0.68} & \textbf{1.79} \\
LSTM-ARFIMA (1) & 17.59\% & 37.88\% & 65.81\% & 0.46 & 0.12 & 0.86 \\
LSTM-ARIMA (2) & 4.29\% & 33.46\% & 66.32\% & 0.13 & 0.01 & 0.24 \\
LSTM-ARFIMA (2) & 16.72\% & 33.19\% & 66.89\% & 0.50 & 0.13 & 0.98 \\
\bottomrule
\multicolumn{7}{p{12cm}}{\raggedright\scriptsize\textit{Note: All values refer to performance indicators derived from predictive models out-of-sample forecasts. The first row represents the benchmark Buy\&Hold strategy. Annotation (1) denotes the hybridization technique of inputting the prediction of the econometric model to the machine learning model, while (2) indicates the hybrid methodology by Zhang (2003).}}
\end{tabular}
\end{center}
\end{table}

\section{Conclusions}

The research aimed to empirically investigate and compare the applicability of various individual and hybrid models in financial asset prediction. The set of individual models was represented by two distinct categories of predictive techniques. The standard ARIMA model and long-memory ARFIMA alternative were applied as the chosen variants of econometric models capable of extracting linear patterns from the data. On the other hand, nonlinear machine learning approaches were implemented using the SVM, XGBoost, and LSTM models, each characterized by distinct features and capabilities. The hybridization process was performed using two techniques: (1) inputting the prediction of the econometric model as an additional explanatory feature for the machine learning model, (2) an additive structure of modeling the econometric model residuals by the machine learning model (Zhang, 2003). As a result, a number of hybrid model architectures were constructed, enabling a thorough analysis of the most suitable individual techniques and the method of combination.

The designed framework was applied to two distinct asset classes: stock market index - S\&P 500, and cryptocurrency - Bitcoin. The data period for S\&P 500 spanned between 1 January 2002 and 31 December 2023. For Bitcoin, the data covered daily observations between 1 January 2015 and 31 December 2023. To generate input for the predictive models, the closing price series was transformed into the logarithmic returns. The process of training and hyperparameter tuning on the in-sample period and forecasting on out-of-sample data was conducted using the novel dynamic sliding window cross-validation technique inspired by Choi et al. (2024). The effectiveness of the models was evaluated based on both prediction error metrics and trading performance indicators. The formulated algorithmic investment strategies were built on \textit{Long-Short} and \textit{Long Only} trading signals with transaction costs included. This approach facilitated simulating the functioning of the market during the backtesting procedure. Moreover, in addition to trading individual assets, a combined approach based on the portfolio of S\&P 500 and Bitcoin was tested.

The conducted empirical investigation provided answers to the earlier stated research questions:

\begin{itemize}
\item RQ1: \textit{Does the use of hybrid models yield more accurate predictions compared to individual linear models and machine learning techniques?}

Based on the results, the answer to this question is not straightforward. For the S\&P 500 index, four hybrid models managed to generate more accurate predictions than their individual components: SVM-ARIMA (1), SVM-ARFIMA (1), LSTM-ARIMA (1), and LSTM-ARFIMA (1). So, the successful implementation of hybridization was dependent on the adopted models. Particularly, the XGBoost-based hybrid models failed to improve the accuracy. Additionally, the method of model combining appears to be a major factor influencing the results. The hybridization approach, which does not assume an additive structure between linear and nonlinear components, contributes to enhanced forecasting performance. On the other hand, ARIMA demonstrated the highest accuracy in modeling the Bitcoin data. Across the hybrid models, the SVM-ARIMA (1) proved to be superior, however, it did not outperform its single components.

\item RQ2: \textit{Is it possible to create a profitable trading strategy (compared to a buy-and-hold benchmark) using the forecasts of constructed hybrid models?}

The risk-adjusted return metrics indicate that the best performance was achieved by the following techniques: SVM - S\&P 500 and \textit{Long-Short}, LSTM-ARFIMA (2) - S\&P 500 and \textit{Long Only}, LSTM-ARIMA (1) - Bitcoin and \textit{Long-Short}, LSTM-ARIMA (1) - Bitcoin and \textit{Long Only}, LSTM-ARIMA (1) - portfolio and \textit{Long-Short}, and LSTM-ARIMA (1) - portfolio and \textit{Long Only}. Many hybrid techniques managed to systematically outperform their single components and Buy\&Hold benchmark. Particularly strong trading performance was delivered by the SVM-ARIMA (1) and LSTM-ARIMA (1) models. So, it can be concluded that appropriately constructed hybrid architecture might lead to the formation of a profitable trading strategy compared to other benchmarks.

\item RQ3: \textit{Which econometric model is better suited for hybrid models in describing linear dependencies in financial markets: the Autoregressive Integrated Moving Average (ARIMA) model or the Autoregressive Fractionally Integrated Moving Average (ARFIMA) model?}

In most cases, the effectiveness of hybrid models implementing ARIMA as the econometric component turned out to be higher in comparison to the ARFIMA-based hybrid techniques. This observation may indicate the lack of long memory for the analyzed assets during the research period. Moreover, the comparison of econometric techniques' individual performance shows the superiority of the ARIMA model, especially in trading applications. Consequently, there are reasons to imply that the ARIMA model provided a more suitable approach for describing linear dependencies in financial markets within the proposed hybrid architectures.

\item RQ4: \textit{Which machine learning or deep learning model is better suited for hybrid models in describing non-linear dependencies in financial markets: Long Short-Term Memory (LSTM), eXtreme Gradient Boosting (XGBoost), or Support Vector Machines (SVM)?}

Two models proved to be well-suited for the proposed hybridization framework: the machine learning SVM model and the deep learning LSTM recurrent neural network. Their superiority was reflected both in the forecasting accuracy and in the automated investment strategies. These two data-based techniques appeared to constitute the appropriate alternatives in modeling nonlinear components in the applied time series data. Significantly inferior performance was recorded by the XGBoost-based hybrid models. This finding is consistent with other studies suggesting poor outcomes for XGBoost in the financial market forecasting (Lv et al., 2019; Chlebus et al., 2021; Mills et al., 2024).

\item RQ5: \textit{Does the selection of hybridization technique influence the results?}

The selection of the hybridization technique was recognized as the major factor influencing the quality of results. The methodology by Zhang (2003), in most cases, failed to deliver satisfactory results. So, it can be stated that the assumption of additive structure between linear and nonlinear components of the time series was not reflected in the applied datasets. On the other hand, the hybridization technique of utilizing the prediction of the econometric model as an additional explanatory variable for the machine learning model was characterized by promising outcomes. In particular, this method achieved superior results in the practical application of the forecasts as trading signals for the constructed investment strategies.

\item RQ6: \textit{Does the selection of the best hybrid model depend on the financial asset class?}

The optimal hybrid model varied according to the category of the financial asset. However, relatively strong performance was recorded for both S\&P 500 and Bitcoin with the SVM-ARIMA (1) and LSTM-ARIMA (1) hybrid models. Additionally, this finding was also emphasized for the portfolio of assets. Therefore, the selection of the best-performing hybridization technique, depending on the type of financial instrument, demonstrated some level of robustness.

\end{itemize}

Summarizing the achieved findings, the hybrid techniques may improve the forecasting and trading performance of the econometric and machine learning models. However, the accomplishment of this outcome is greatly dependent on the selection of the applied individual models and the method of their combination. The applicability of the particular hybrid architectures should be thoroughly evaluated during the cross-validation and backtesting procedures before employing them in real market applications. This study proposes a comprehensive framework for achieving this goal, which may prove useful to both researchers and market practitioners.

In the end, the study has faced some limitations that should be addressed in future research. First of all, the constructed models were only applied to a limited number of an arbitrarily selected class of assets. Potential extensions of the study include the implementation of the proposed hybrid methodology to other categories of financial assets, including commodities, currencies, and individual stocks. In addition, instead of modeling individual assets, attention should be put on applying the presented framework to a portfolio of multiple financial instruments. Another direction of future empirical investigation might include the incorporation of high-frequency data. This approach would allow for testing the hybridization in a more dynamic trading environment. Lastly, additional representatives of econometric and machine learning models should be employed as individual components of the hybrid framework. Furthermore, alternative forms of hybridization could be evaluated, for example, the use of genetic algorithms (Wang et al., 2012; Rather et al., 2015).

\section*{References}

\begin{enumerate}
    \item Akyildirim, E., Cepni, O., Corbet, S., \& Uddin, G. S. (2023). Forecasting mid-price movement of Bitcoin futures using machine learning. \textit{Annals of Operations Research}, \textit{330}(1), 553-584.
    \item Al-Selwi, S. M., Hassan, M. F., Abdulkadir, S. J., Muneer, A., Sumiea, E. H., Alqushaibi, A., \& Ragab, M. G. (2024). RNN-LSTM: From applications to modeling techniques and beyond—Systematic review. \textit{Journal of King Saud University-Computer and Information Sciences}, 102068.
    \item Aladag, C. H., Egrioglu, E., \& Kadilar, C. (2012). Improvement in forecasting accuracy using the hybrid model of ARFIMA and feed forward neural network. \textit{American Journal of Intelligent Systems}, \textit{2}(2), 12-17.
    \item Assaf, A. (2006). Dependence and mean reversion in stock prices: The case of the MENA region. \textit{Research in International Business and Finance}, \textit{20}(3), 286-304.
    \item Atsalakis, G. S., \& Valavanis, K. P. (2009). Surveying stock market forecasting techniques–Part II: Soft computing methods. \textit{Expert Systems with Applications}, \textit{36}(3), 5932-5941.
    \item Bailey, D. H., Borwein, J. M., De Prado, M. L., \& Zhu, Q. J. (2014). Pseudomathematics and financial charlatanism: The effects of backtest overfitting on out-of-sample performance. \textit{Notices of the AMS}, \textit{61}(5), 458-471.
    \item Bailey, D. H., Borwein, J. M., De Prado, M. L., Salehipour, A., \& Zhu, Q. J. (2016). Backtest overfitting in financial markets. \textit{Automated Trader}.
    \item Bao, W., Yue, J., \& Rao, Y. (2017). A deep learning framework for financial time series using stacked autoencoders and \textit{Long-Short} term memory. \textit{PloS one}, \textit{12}(7), e0180944.
    \item Barkoulas, J. T., Baum, C. F., \& Travlos, N. (2000). Long memory in the Greek stock market. \textit{Applied Financial Economics}, \textit{10}(2), 177-184.
    \item Bennett, K. P., \& Campbell, C. (2000). Support vector machines: hype or hallelujah?. \textit{ACM SIGKDD explorations newsletter}, \textit{2}(2), 1-13.
    \item Bhardwaj, G., \& Swanson, N. R. (2006). An empirical investigation of the usefulness of ARFIMA models for predicting macroeconomic and financial time series. \textit{Journal of econometrics}, \textit{131}(1-2), 539-578.
    \item Bieganowski, B., \& Slepaczuk, R. (2024). Supervised Autoencoder MLP for Financial Time Series Forecasting. \textit{arXiv preprint arXiv:2404.01866}.
    \item Bukhari, A. H., Raja, M. A. Z., Sulaiman, M., Islam, S., Shoaib, M., \& Kumam, P. (2020). Fractional neuro-sequential ARFIMA-LSTM for financial market forecasting. \textit{Ieee Access}, \textit{8}, 71326-71338.
    \item Bouteska, A., Abedin, M. Z., Hajek, P., \& Yuan, K. (2024). Cryptocurrency price forecasting–a comparative analysis of ensemble learning and deep learning methods. \textit{International Review of Financial Analysis}, \textit{92}, 103055.
    \item Box, G.E.P.; Jenkins, G.M. \textit{Time Series Analysis Forecasting and Control}. Holden Day: San Francisco, CA, USA, 1976.
    \item Bustos, O., \& Pomares-Quimbaya, A. (2020). Stock market movement forecast: A systematic review. \textit{Expert Systems with Applications}, \textit{156}, 113464.
    \item Cao, L., \& Tay, F. E. (2001). Financial forecasting using support vector machines. \textit{Neural Computing \& Applications}, \textit{10}, 184-192.
    \item Chaâbane, N. (2014). A hybrid ARFIMA and neural network model for electricity price prediction. \textit{International journal of electrical power \& energy systems}, \textit{55}, 187-194.
    \item Chatzis, S. P., Siakoulis, V., Petropoulos, A., Stavroulakis, E., \& Vlachogiannakis, N. (2018). Forecasting stock market crisis events using deep and statistical machine learning techniques. \textit{Expert systems with applications}, \textit{112}, 353-371.
    \item Chen, T., \& Guestrin, C. (2016, August). Xgboost: A scalable tree boosting system. \textit{In Proceedings of the 22nd acm sigkdd international conference on knowledge discovery and data mining} (pp. 785-794).
    \item Cheung, Y. W., \& Lai, K. S. (1995). A search for long memory in international stock market returns. \textit{Journal of International Money and Finance}, \textit{14}(4), 597-615.
    \item Chlebus, M., Dyczko, M., \& Woźniak, M. (2021). Nvidia's stock returns prediction using machine learning techniques for time series forecasting problem. \textit{Central European Economic Journal}, \textit{8}(55).
    \item Choi, W., Jang, S., Kim, S., Park, C., Park, S., \& Song, S. (2024). Return prediction by machine learning for the Korean stock market. \textit{Journal of the Korean Statistical Society}, \textit{53}(1), 248-280.
    \item Cocco, L., Tonelli, R., \& Marchesi, M. (2021). Predictions of bitcoin prices through machine learning based frameworks. \textit{PeerJ Computer Science}, \textit{7}, e413.
    \item Cortes, C., \& Vapnik, V. (1995). Support-vector networks. \textit{Machine learning}, \textit{20}, 273-297.
    \item De Gooijer, J. G., \& Hyndman, R. J. (2006). 25 years of time series forecasting. \textit{International Journal of Forecasting}, \textit{22}(3), 443-473.
    \item De Prado, M. L. (2015). The future of empirical finance. \textit{Journal of Portfolio Management}, \textit{41}(4).
    \item De Prado, M. L. (2018). \textit{Advances in financial machine learning}. John Wiley \& Sons.
    \item De Prado, M. L. (2019). Beyond econometrics: A roadmap towards financial machine learning. \textit{Available at SSRN 3365282}.
    \item Dudek, G., Fiszeder, P., Kobus, P., \& Orzeszko, W. (2024). Forecasting cryptocurrencies volatility using statistical and machine learning methods: A comparative study. \textit{Applied Soft Computing}, \textit{151}, 111132.
    \item Fama, E. F. (1970). Efficient capital markets. \textit{Journal of finance}, \textit{25}(2), 383-417.
    \item Fischer, T., \& Krauss, C. (2018). Deep learning with long short-term memory networks for financial market predictions. \textit{European journal of operational research}, \textit{270}(2), 654-669.
    \item Floros, C., Jaffry, S., \& Valle Lima, G. (2007). Long memory in the Portuguese stock market. \textit{Studies in Economics and Finance}, \textit{24}(3), 220-232.
    \item Freund, Y., \& Schapire, R. E. (1997). A decision-theoretic generalization of on-line learning and an application to boosting. \textit{Journal of computer and system sciences}, \textit{55}(1), 119-139.
    \item Gómez, S. C., \& Ślepaczuk, R. (2021). \textit{Robust optimisation in algorithmic investment strategies} (No. 2021-27).
    \item Geboers, H., Depaire, B., \& Annaert, J. (2023). A review on drawdown risk measures and their implications for risk management. \textit{Journal of Economic Surveys}, 37(\textit{3}), 865-889.
    \item Gers, F. A., Schmidhuber, J., \& Cummins, F. (2000). Learning to forget: Continual prediction with LSTM. \textit{Neural computation}, \textit{12}(10), 2451-2471.
    \item Gers, F. A., Schraudolph, N. N., \& Schmidhuber, J. (2002). Learning precise timing with LSTM recurrent networks. \textit{Journal of machine learning research}, \textit{3}(Aug), 115-143.
    \item Granger, C. W., \& Ding, Z. (1996). Varieties of long memory models. \textit{Journal of econometrics}, \textit{73}(1), 61-77.
    \item Granger, C. W., \& Joyeux, R. (1980). An introduction to long‐memory time series models and fractional differencing. \textit{Journal of time series analysis}, \textit{1}(1), 15-29.
    \item Granger, C. W., \& Newbold, P. (1974). Spurious regressions in econometrics. \textit{Journal of econometrics}, \textit{2}(2), 111-120.
    \item Greff, K., Srivastava, R. K., Koutník, J., Steunebrink, B. R., \& Schmidhuber, J. (2016). LSTM: A search space odyssey. \textit{IEEE transactions on neural networks and learning systems}, \textit{28}(10), 2222-2232.
    \item Grudniewicz, J., \& Ślepaczuk, R. (2023). Application of machine learning in algorithmic investment strategies on global stock markets. \textit{Research in International Business and Finance}, \textit{66}, 102052
    \item Harvey, A. C. (1990). Forecasting, structural time series models and the Kalman filter.
    \item Hibon, M., \& Evgeniou, T. (2005). To combine or not to combine: selecting among forecasts and their combinations. \textit{International journal of forecasting}, \textit{21}(1), 15-24.
    \item Hochreiter, S., \& Schmidhuber, J. (1997). Long short-term memory. \textit{Neural computation}, \textit{9}(8), 1735-1780.
    \item Hochreiter, S. (1998). The vanishing gradient problem during learning recurrent neural nets and problem solutions. \textit{International Journal of Uncertainty, Fuzziness and Knowledge-Based Systems}, \textit{6}(02), 107-116.
    \item Hosking, J. (1981). Fractional differencing. \textit{Biometrika 68}(1), 165–175
    \item Hsieh, T. J., Hsiao, H. F., \& Yeh, W. C. (2011). Forecasting stock markets using wavelet transforms and recurrent neural networks: An integrated system based on artificial bee colony algorithm. \textit{Applied Soft Computing}, \textit{11}(2), 2510-2525.
    \item Hsu, M. W., Lessmann, S., Sung, M. C., Ma, T., \& Johnson, J. E. (2016). Bridging the divide in financial market forecasting: machine learners vs. financial economists. \textit{Expert Systems with Applications}, \textit{61}, 215-234.
    \item Huang, W., Nakamori, Y., \& Wang, S. Y. (2005). Forecasting stock market movement direction with support vector machine. \textit{Computers \& operations research}, \textit{32}(10), 2513-2522.
    \item Hudson, R. S., \& Gregoriou, A. (2015). Calculating and comparing security returns is harder than you think: A comparison between logarithmic and simple returns. \textit{International Review of Financial Analysis}, \textit{38}, 151-162.
    \item Ince, H., \& Trafalis, T. B. (2008). Short term forecasting with support vector machines and application to stock price prediction. \textit{International Journal of General Systems}, \textit{37}(6), 677-687.
    \item Jabeur, S. B., Mefteh-Wali, S., \& Viviani, J. L. (2024). Forecasting gold price with the XGBoost algorithm and SHAP interaction values. \textit{Annals of Operations Research}, \textit{334}(1), 679-699.
    \item Jiang, Y., Nie, H., \& Ruan, W. (2018). Time-varying long-term memory in Bitcoin market. \textit{Finance Research Letters}, \textit{25}, 280-284.
    \item Kashif, K., \& Ślepaczuk, R. (2025). LSTM-ARIMA as a hybrid approach in algorithmic investment strategies. \textit{Knowledge-Based Systems}.
    \item Kim, K. J. (2003). Financial time series forecasting using support vector machines. \textit{Neurocomputing}, \textit{55}(1-2), 307-319.
    \item Kobiela, D., Krefta, D., Król, W., \& Weichbroth, P. (2022). ARIMA vs LSTM on NASDAQ stock exchange data. \textit{Procedia Computer Science}, \textit{207}, 3836-3845.
    \item Kolbadi, P., \& Ahmadinia, H. (2011). Examining Sharp, Sortino and Sterling ratios in portfolio management, evidence from Tehran stock exchange. \textit{International Journal of Business and Management}, \textit{6}(4), 222.
    \item Kosc, K., Sakowski, P., \& Ślepaczuk, R. (2019). Momentum and contrarian effects on the cryptocurrency market. \textit{Physica A: Statistical Mechanics and its Applications}, \textit{523}, 691-701.
    \item Koustas, Z., \& Serletis, A. (2005). Rational bubbles or persistent deviations from market fundamentals?. \textit{Journal of Banking \& Finance}, \textit{29}(10), 2523-2539.
    \item Kumar, M. (2010). Modelling exchange rate returns using non-linear models. \textit{Margin: The Journal of Applied Economic Research}, \textit{4}(1), 101-125.
    \item Kumar, M., \& Thenmozhi, M. (2014). Forecasting stock index returns using ARIMA-SVM, ARIMA-ANN, and ARIMA-random forest hybrid models. \textit{International Journal of Banking, Accounting and Finance}, \textit{5}(3), 284-308.
    \item Kwon, D. H., Kim, J. B., Heo, J. S., Kim, C. M., \& Han, Y. H. (2019). Time series classification of cryptocurrency price trend based on a recurrent LSTM neural network. \textit{Journal of Information Processing Systems}, \textit{15}(3), 694-706.
    \item López-Martín, C., Benito Muela, S., \& Arguedas, R. (2021). Efficiency in cryptocurrency markets: New evidence. \textit{Eurasian Economic Review},  \textit{11}(3), 403-431.
    \item Li, P., \& Zhang, J. S. (2018). A new hybrid method for China’s energy supply security forecasting based on ARIMA and XGBoost. \textit{Energies}, \textit{11}(7), 1687.
    \item Liu, J., \& Serletis, A. (2019). Volatility in the cryptocurrency market. \textit{Open Economies Review}, \textit{30}(4), 779-811.
    \item Kijewski, M., Ślepaczuk, R., \& Wysocki, M. (2024). Predicting prices of S\&P 500 index using classical methods and recurrent neural networks.
    \item Khursheed, A., Naeem, M., Ahmed, S., \& Mustafa, F. (2020). Adaptive market hypothesis: An empirical analysis of time–varying market efficiency of cryptocurrencies. \ textit{Cogent Economics \& Finance},\ textit{8}(1), 1719574.
    \item Lo, A. W. (1991). Long-term memory in stock market prices. \textit{Econometrica: Journal of the Econometric Society}, 1279-1313.
    \item Lv, D., Yuan, S., Li, M., \& Xiang, Y. (2019). An empirical study of machine learning algorithms for stock daily trading strategy. \textit{Mathematical Problems in Engineering}, \textit{2019}(1), 7816154.
    \item Ma, Y., Han, R., \& Wang, W. (2020). Prediction-based portfolio optimization models using deep neural networks. \textit{Ieee Access}, \textit{8}, 115393-115405.
    \item Ma, Y., Han, R., \& Wang, W. (2021). Portfolio optimization with return prediction using deep learning and machine learning. \textit{Expert Systems with Applications}, \textit{165}, 113973.
    \item Malandri, L., Xing, F. Z., Orsenigo, C., Vercellis, C., \& Cambria, E. (2018). Public mood–driven asset allocation: The importance of financial sentiment in portfolio management. \textit{Cognitive Computation}, \textit{10}(6), 1167-1176.
    \item Malkiel, B. G. (2003). The efficient market hypothesis and its critics. \textit{Journal of Economic Perspectives}, \textit{17}(1), 59-82.
    \item MathWorks. (2025, April 27). \textit{Understanding support vector machine regression.} MathWorks. \href{https://www.mathworks.com/help/stats/understanding-support-vector-machine-regression.html}{mathworks.com/.../svm-regression}
    \item McNally, S., Roche, J., \& Caton, S. (2018, March). Predicting the price of bitcoin using machine learning. In \textit{2018 26th euromicro international conference on parallel, distributed and network-based processing (PDP)} (pp. 339-343). IEEE.
    \item Michańków, J., Sakowski, P., \& Ślepaczuk, R. (2022). LSTM in algorithmic investment strategies on BTC and S\&P500 index. \textit{Sensors}, \textit{22}(3), 917.
    \item Mills, E. F. E. A., Liao, Y., \& Deng, Z. (2024). Data-driven price trends prediction of Ethereum: A hybrid machine learning and signal processing approach. \textit{Blockchain: Research and Applications}, \textit{5}(4), 100231.
    \item Newbold, P. (1975). The principles of the Box-Jenkins approach. \textit{Journal of the Operational Research Society}, \textit{26}(2), 397-412.
    \item Nobre, J., \& Neves, R. F. (2019). Combining principal component analysis, discrete wavelet transform and XGBoost to trade in the financial markets. \textit{Expert Systems with Applications}, \textit{125}, 181-194.
    \item Pai, P. F., \& Lin, C. S. (2005). A hybrid ARIMA and support vector machines model in stock price forecasting. \textit{Omega}, \textit{33}(6), 497-505.
    \item Pérez-Pons, M. E., Parra-Dominguez, J., Omatu, S., Herrera-Viedma, E., \& Corchado, J. M. (2022). Machine learning and traditional econometric models: a systematic mapping study. \textit{Journal of Artificial Intelligence and Soft Computing Research}, \textit{12}.
    \item Rather, A. M., Agarwal, A., \& Sastry, V. N. (2015). Recurrent neural network and a hybrid model for prediction of stock returns. \textit{Expert Systems with Applications}, \textit{42}(6), 3234-3241.
    \item Rouf, N., Malik, M. B., Arif, T., Sharma, S., Singh, S., Aich, S., \& Kim, H. C. (2021). Stock market prediction using machine learning techniques: a decade survey on methodologies, recent developments, and future directions. \textit{Electronics}, \textit{10}(21), 2717.
    \item Ryś, P., \& Ślepaczuk, R. (2019). Machine Learning Methods in Algorithmic Trading Strategy Optimization–Design and Time Efficiency. \textit{Central European Economic Journal}, \textit{5}(52).
    \item Schapire, R. E. (1999, July). A brief introduction to boosting. In \textit{Ijcai} (Vol. 99, No. 999, pp. 1401-1406).
    \item Shah, D., Isah, H., \& Zulkernine, F. (2019). Stock market analysis: A review and taxonomy of prediction techniques. \textit{International Journal of Financial Studies}, \textit{7}(2), 26.
    \item Sharpe, W. F. (1966). Mutual fund performance. \textit{The Journal of business}, \textit{39}(1), 119-138.
    \item Shen, S., Jiang, H., \& Zhang, T. (2012). Stock market forecasting using machine learning algorithms. \textit{Department of Electrical Engineering, Stanford University, Stanford, CA}, 1-5.
    \item Stoica, P., \& Selen, Y. (2004). Model-order selection: a review of information criterion rules. \textit{IEEE Signal Processing Magazine}, \textit{21}(4), 36-47.
    \item Taskaya-Temizel, T., \& Casey, M. C. (2005). A comparative study of autoregressive neural network hybrids. \textit{Neural Networks}, \textit{18}(5-6), 781-789.
    \item Terui, N., \& Van Dijk, H. K. (2002). Combined forecasts from linear and nonlinear time series models. \textit{International Journal of Forecasting}, \textit{18}(3), 421-438.
    \item Viéitez, A., Santos, M., \& Naranjo, R. (2024). Machine learning Ethereum cryptocurrency prediction and knowledge-based investment strategies. \\ \textit{Knowledge-Based Systems}, \textit{299}, 112088.
    \item Vo, N., \& Ślepaczuk, R. (2022). Applying hybrid ARIMA-SGARCH in algorithmic investment strategies on S\&P500 index. \textit{Entropy}, \textit{24}(2), 158.
    \item Wang, J. J., Wang, J. Z., Zhang, Z. G., \& Guo, S. P. (2012). Stock index forecasting based on a hybrid model. \textit{Omega}, \textit{40}(6), 758-766.
    \item Weng, B., Ahmed, M. A., \& Megahed, F. M. (2017). Stock market one-day ahead movement prediction using disparate data sources. \textit{Expert Systems with Applications}, \textit{79}, 153-163.
    \item Wu, X., Wu, L., \& Chen, S. (2022). Long memory and efficiency of Bitcoin during COVID-19. \textit{Applied Economics}, \textit{54}(4), 375-389.
    \item Yang, C., Zhai, J., \& Tao, G. (2020). Deep learning for price movement prediction using convolutional neural network and long short‐term memory. \textit{Mathematical Problems in Engineering}, \textit{2020}(1), 2746845.
    \item Zhang, G. P. (2003). Time series forecasting using a hybrid ARIMA and neural network model. \textit{Neurocomputing}, \textit{50}, 159-175.
\end{enumerate}

\end{document}